# *Theory of Quantum Friction*

*Mário G. Silveirinha*[*]

[(1)]*University of Coimbra, Department of Electrical Engineering – Instituto de Telecomunicações, Portugal, mario.silveirinha@co.it.pt*

**Abstract**

Here, we develop a comprehensive quantum theory for the phenomenon of quantum friction. Based on a theory of macroscopic quantum electrodynamics for unstable systems, we calculate the quantum expectation of the friction force at zero temperature, and link the friction effect to the emergence of system instabilities related to the Cherenkov effect. These instabilities may occur due to the hybridization of particular guided modes supported by the individual moving bodies, and selection rules for the interacting modes are derived. It is proven that the quantum friction effect can take place even when the interacting bodies are lossless and made of nondispersive dielectrics.

---

[*] To whom correspondence should be addressed: E-mail: *mario.silveirinha@co.it.pt*

# I. Introduction

Quantum friction is a theory that predicts that two uncharged polarizable bodies moving relative to each other experience a force of quantum origin that tends to work against the relative motion [1-8]. This effect is predicted to take place even a zero temperature and when the surfaces of the moving bodies are flat and perfectly smooth (the materials are regarded continuous media). A physical picture is that the electric dipoles created by the quantum fluctuations in one of the surfaces induce image electric dipoles on the other surface, which, when the bodies are in relative parallel motion, lag behind and originate a van-der-Waals type attraction [1]. The quantum friction theory is not consensual and has been recently debated [4, 9, 10]. Quantum friction has also been studied in rotating dielectric bodies [11-13].

For many authors, quantum friction is understood as a purely quantum effect with no classical analogue. Recently, it was shown in Ref. [8] that for the case of sliding monoatomic surfaces the effect of friction is associated with electromagnetic instabilities in moving media that can lead to the creation of polaritons. Very interestingly, these electromagnetic instabilities are partly connected to the Cherenkov [14] and Smith-Purcell effects [15], and can be predicted by *classical* electrodynamics. Related electromagnetic instabilities have been discussed in the context of plasma physics, with application in the design of terahertz traveling-wave oscillators and amplifiers [16]. These electromagnetic instabilities are manifested in the fact that the system may support natural modes of oscillation that *grow exponentially* with time [8, 16, 17], even in presence of material loss [18]. In this article, we establish the definite and missing link between this classical effect and quantum friction.

Most of the available theories of quantum friction are based either on semi-classical arguments or on first order perturbation quantum theory. In Ref. [17], we developed (in the framework of macroscopic quantum electrodynamics) a theory for the quantization of the electromagnetic field in moving media systems with electromagnetic instabilities. Using this formalism, here we derive the quantum friction force from first principles at zero temperature, and prove that it has a *dynamical character*, i.e. the expectation of the friction force varies with time. We prove that in the "pseudo-ground" state of the system [17], i.e. in the state wherein the oscillations of the quantum fields in the two moving bodies are minimal, the expectation of the friction force vanishes. However, as time passes the friction force builds up exponentially, as long as the velocity of the bodies is enforced to be constant through the application of an external force. Interestingly, we establish a precise connection between our theory and the semiclassical theory of Pendry [1, 4]. We prove that Pendry's friction force corresponds to our dynamic friction force calculated at the time instant wherein the first "excitation" is generated.

The usual explanation found in the literature for quantum friction is related to material loss, such that the frictional work done on a given body is dissipated in the electrical resistance of the dielectric [3-5]. For example, in Ref. [8] it was found that the friction force vanishes in the lossless limit wherein the material responds instantaneously to the local fields. However, the analysis of Ref. [8] ignores the retardation effects due to the finite speed of light ($c \to \infty$), and thus in the absence of material dispersion the interaction between different electric dipoles is effectively instantaneous, and, moreover, it is impossible to surpass the Cherenkov critical velocity when $c \to \infty$. Here, we prove that when wave retardation is properly taken into account it is possible to have a frictional force even when the local material response is instantaneous. This demonstrates that the friction effect does *not* require material loss. Thus, surprisingly, the friction force can be nonzero even when the interacting bodies are made of nondispersive lossless dielectrics. In a recent work [19], Maghrebi *et al*, have independently demonstrated (based on a scalar field theory) a connection between noncontact friction and Cherenkov radiation, consistent with our studies [8, 17] (see also Ref. [20]). The emergence of electromagnetic instabilities above the threshold velocity for quantum Cherenkov emission was however not discussed by the authors of Ref. [19].

The article is organized as follows. In Sect. II, we review and extend the formalism developed in Ref. [17], which is the basis of our theory. In Sect. III, the friction force quantum operator is derived. In Sect. IV, the selection rules for guided modes that originate system instabilities are obtained. These selection rules complement the findings of Ref. [17]. In Sect. V we compute the quantum expectation of the friction force, and Sect. VI reports several numerical examples and an explicit comparison with the theory of Pendry. The conclusion is drawn in Sect. VII.

## II. Waves in Moving Media

### A. *Material bodies coupled by the electromagnetic field*

We are interested in the dynamics of a set of rigid lossless non-magnetic material bodies ($i$=1,2,…) coupled by the electromagnetic field. Let $\mathbf{v}_i$ be the velocity of the $i$-th body center of mass. Consistent with our previous work [17], it is shown in Appendix A that provided $v_i/c \ll 1$ the total energy ($H_{tot}$) of the system can be written as,

$$H_{tot} = \sum_i \mathbf{v}_i \cdot \left( \mathbf{p}_{can,i} - \frac{M_i \mathbf{v}_i}{2} \right) + H_{EM,P}, \qquad (1)$$

where $M_i$ is the mass of the $i$-th body, $\mathbf{p}_{can,i}$ is the total canonical momentum of the $i$-th body, and $H_{EM,P}$ is by definition the "wave energy":

$$H_{EM,P} = \frac{1}{2} \int d^3\mathbf{r}\, \mathbf{B} \cdot \mathbf{H} + \mathbf{D} \cdot \mathbf{E} \;. \qquad (2)$$

As is well known, for charged particles the canonical momentum ($\mathbf{p}_{can}$) differs from the kinetic momentum ($\mathbf{p}_{kin} = M\mathbf{v}$). The canonical momentum is the conjugate quantity to the position vector [21]. For relatively weak field amplitudes the general term of the sum in Eq. (1) is approximately equal to $p_{can,i}^2 / 2M_i$. We show explicitly in Appendix A that the total energy is always *nonnegative*, $H_{tot} \geq 0$.

From Eq. (1), it is seen that the total energy has a wave part ($H_{EM,P}$), as well as a part related to the canonical momentum of the moving bodies. The wave energy ($H_{EM,P}$) is not purely electromagnetic and includes also part of the energy stored in matter (e.g. the energy associated with dipole vibrations and part of the energy associated with the translational motion) [17]. It is proven in Appendix A that our system satisfies exactly the following conservation law:

$$\frac{dH_{tot}}{dt} = \sum_i \mathbf{F}_{tot,i}^{ext} \cdot \mathbf{v}_i - \int \mathbf{E} \cdot \mathbf{j}_{ext} d^3\mathbf{r}. \qquad (3)$$

where $\mathbf{j}_{ext}$ is an hypothetical external electric current density (in this work $\mathbf{j}_{ext} = 0$), and $\mathbf{F}_{tot,i}^{ext}$ is the external force acting on the $i$-th moving body. For a closed system $\mathbf{j}_{ext} = 0$ and $\mathbf{F}_{tot,i}^{ext} = 0$, and hence in that case $H_{tot}$ is conserved, even in presence of the wave instabilities discussed ahead.

In Appendix A (see also Ref. [17]), it is demonstrated that when the considered bodies are invariant to translations along the *x*-direction, the time derivative of the *x*-component of the total momentum $p_i = p_{kin,i} + p_{EM,i}$ associated with the *i-th* body satisfies:

$$\frac{dp_i}{dt} \equiv \frac{dp_{kin,i}}{dt} + \frac{dp_{EM,i}}{dt} = \frac{dp_{w,i}}{dt} + F_{i,x}^{ext}. \tag{4}$$

In the above, $F_{i,x}^{ext}$ represents the *x*-component of $\mathbf{F}_{tot,i}^{ext}$, $p_{kin,i}$ is the *x*-component of the kinetic momentum of the *i-th* body, and the electromagnetic momentum ($p_{EM,i}$) and the wave momentum ($p_{w,i}$) are defined by

$$p_{EM,i} = \frac{1}{c^2} \int_{V_i} (\mathbf{E} \times \mathbf{H}) \cdot \hat{\mathbf{x}} \, d^3\mathbf{r}, \tag{5a}$$

$$p_{w,i} = \int_{V_i} (\mathbf{D} \times \mathbf{B}) \cdot \hat{\mathbf{x}} \, d^3\mathbf{r}, \tag{5b}$$

where $V_i$ is the volume of the pertinent body. Furthermore, the time derivative of the canonical momentum equals the external force (see Appendix A):

$$dp_{can,i} / dt = F_{i,x}^{ext}. \tag{6}$$

Thus, Eq. (4) is compatible with the decompositions for the total momentum $p_i = p_{kin,i} + p_{EM,i} = p_{can,i} + p_{wv,i}$ [22, 23, 24].

### B.  *Material bodies with time independent velocities*

Next, we consider a system of polarizable non-dispersive moving bodies invariant to translations along the *x* and *y* directions (Fig. 1a). It is supposed that the relevant bodies move with a time independent velocity $\mathbf{v} = v(z)\hat{\mathbf{x}}$ along the *x* direction. We allow $\mathbf{v}$ to depend on *z* because different bodies can have different velocities. Each body is characterized by the material parameters $\varepsilon = \varepsilon(z)$ and $\mu = \mu_0$ in the respective co-moving frame. In this work, we are interested in velocities larger than the Cherenkov emission threshold ($|v| > c/n$, with $n(z) = \sqrt{\varepsilon(z)/\varepsilon_0}$ the refractive index in the co-moving frame), and thus it is required that $n \gg 1$ so that the condition $v/c \ll 1$ is observed.

One crucial point is that when the velocities of the moving bodies are time independent the dynamics of the electromagnetic field becomes decoupled from the dynamics of the canonical momenta ($\mathbf{p}_{can,i}$), and hence it can be studied based simply on the Maxwell's equations and on constitutive relations of the polarizable bodies [Eqs. (A1) and (A4)]. Moreover, in such a case the *wave part* of the system energy and momentum is conserved [17]. Indeed, from Eqs. (1), (3) and (6) and $\mathbf{j}_{ext} = 0$ it is seen that when the velocities of the moving bodies are time independent the wave energy satisfies $dH_{EM,P}/dt = 0$.

In particular, in a generic reference frame a moving material is seen as a non-reciprocal bianisotropic medium [25]. Thus, in a fixed reference frame (laboratory frame) the moving bodies are characterized by a 6×6 material matrix $\mathbf{M} = \mathbf{M}(z)$, which relates the classical $\mathbf{G} = (\mathbf{D} \quad \mathbf{B})^T$ fields and the classical $\mathbf{F} = (\mathbf{E} \quad \mathbf{H})^T$ fields as $\mathbf{G} = \mathbf{M} \cdot \mathbf{F}$ [17, 22, 25]. In a relativistic framework, the material matrix $\mathbf{M}(z)$ is written in terms of the material parameters in the co-moving frame ($\varepsilon = \varepsilon(z)$ and $\mu = \mu_0$) and of the velocity ($\mathbf{v} = v(z)\hat{\mathbf{x}}$) as detailed in Ref. [17]. In the absence of radiation sources, the electromagnetic fields satisfy:

$$\hat{N} \cdot \mathbf{F} = i\mathbf{M} \cdot \frac{\partial \mathbf{F}}{\partial t}, \qquad \text{with } \hat{N} = \begin{pmatrix} 0 & i\nabla \times \\ -i\nabla \times & 0 \end{pmatrix}. \tag{7}$$

We suppose that our system is effectively homogeneous for translations along *x* (the slabs are infinitely wide), and thus the spatial-domain is terminated with periodic boundary conditions. Hence, the natural modes of oscillation of the electromagnetic field (i.e. waves with a time dependence $e^{-i\omega t}$) vary as $e^{i\mathbf{k}\cdot\mathbf{r}}$ with *x* and *y*, being $\mathbf{k} = (k_x, k_y, 0)$ a real-valued wave vector, and satisfy:

$$\hat{N} \cdot \mathbf{F}_\omega = \omega \mathbf{M} \cdot \mathbf{F}_\omega. \tag{8}$$

The material matrix $\mathbf{M}$ is symmetric and real-valued. Moreover, the matrix $\mathbf{M}(z)$ is positive definite when $v(z) < c/n(z)$. However, when $v(z) > c/n(z)$, i.e. if a given body has a velocity larger than the Cherenkov emission threshold, the material matrix $\mathbf{M}(z)$ becomes indefinite. The fact that $\mathbf{M}(z)$ can be an indefinite matrix has important implications. Indeed, the stored wave energy (2) can be written in terms of $\mathbf{M}$ as $H_{EM,P} = \langle \mathbf{F} | \mathbf{F} \rangle$ where we put:

$$\langle \mathbf{F}_2 | \mathbf{F}_1 \rangle = \frac{1}{2} \int d^3\mathbf{r} \, \mathbf{F}_2^* \cdot \mathbf{M}(z) \cdot \mathbf{F}_1. \tag{9}$$

Thus, when $\mathbf{M}(z)$ is indefinite the stored wave energy can be *negative* and has no lower bound [17]. Related to this result, in Ref. [7] it was found that the Hamiltonian of a system with parts in relative motion contains negative-energy normal modes. This property can have dramatic consequences. It was proven in Ref. [17], that the interaction between two moving bodies such that the *wave* energies stored in each of the bodies have *opposite signs* can originate system instabilities so that the electromagnetic field may support natural modes of oscillation with $\omega = \omega' + i\lambda$ complex valued. In particular, when $\lambda > 0$ the electromagnetic field oscillations may grow exponentially in time, as long as the velocity of the moving bodies is kept constant [17]. These instabilities take place even though there is no explicit source of excitation and in presence of strong material loss [18].

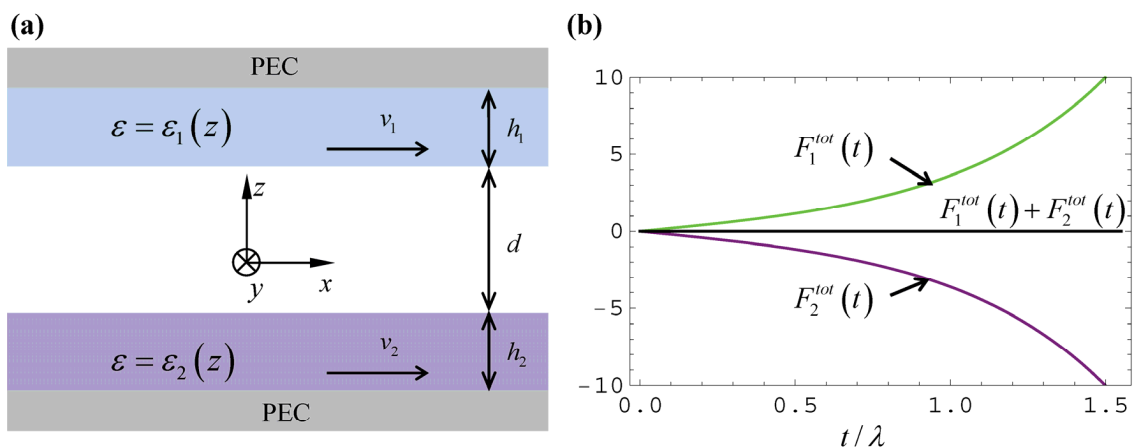

Fig. 1. (Color online) (a) Two dielectric slabs move with different velocities with respect to a certain laboratory frame. The system is invariant to translations along the *x*-direction. In this sketch, it is assumed that the dielectric slabs are backed by perfect electric conductors (PEC). The slabs are made of nondispersive dielectrics and are separated by a vacuum gap. (b) Sketch of the typical time evolution of the (quantum expectation of the) friction force (in arbitrary units) acting on each body when the velocity of the two bodies does not vary appreciably in the considered time interval. The system is in the pseudo-ground state $|\Omega\rangle$ at $t = 0$ and $v_2 > v_1$ in this example. For simplicity, we consider only the contribution of a pair of oscillators associated with the complex valued frequencies $\omega_c = \omega' \pm i\lambda$ so that $F_i^{tot} \sim \sinh(2\lambda t)$ [Eq. (22)].

## C. The force acting on a moving slab

As already discussed in Ref. [17], the wave instabilities imply the emergence of a friction force associated with the radiation drag whose effect is to act against the relative motion of the bodies. For closed systems ($\mathbf{j}_{ext} = 0$ and $\mathbf{F}^{ext}_{tot,i} = 0$), this feedback mechanism results in a decrease of the relative velocity of the bodies, and this ultimately prevents the continued exponential growth of the fields. From Eq. (4) one sees that in the absence of an external force the time rate of change of total momentum (i.e. the sum of the matter and electromagnetic momenta) enclosed in the *i*-th slab is $dp_{wv,i}/dt$. Thus, $dp_{wv,i}/dt$ is a stress associated with the wave flow:

$$F_i^{tot} = \frac{dp_{w,i}}{dt}. \tag{10}$$

It will be seen that in the quantum vacuum this stress acts against the relative motion, and will be responsible by a *friction force*. Thus, in the absence of an external force the velocity of the moving bodies typically changes with time.

Thus, the velocities can remain constant – as will be assumed in this article – only at the expense of applying an external action that counterbalances the friction force. It is seen from Eq. (4) that the external force required to maintain the velocity of the *i*-th body constant ($dp_{kin,i}/dt = 0$) is given by $F_{i,x}^{ext} = -dp_{ps,i}/dt$, being $p_{ps,i} = p_{wv,i} - p_{EM,i}$ the *x*-component of the pseudo-momentum of the *i*-th slab. Thus, $dp_{ps,i}/dt$ is the friction force acting on the matter enclosed in the *i*-th slab:

$$F_i^{mat} = \frac{dp_{ps,i}}{dt}, \quad \text{with} \quad p_{ps,i} = \int_{V_i} \left( \mathbf{D} \times \mathbf{B} - \frac{1}{c^2} \mathbf{E} \times \mathbf{H} \right) \cdot \hat{\mathbf{x}} \, d^3\mathbf{r}. \tag{11}$$

Evidently, in general the two forces are different $F_i^{tot} \neq F_i^{mat}$, but it be seen later that $F_i^{mat} \approx F_i^{tot}$. Note that $F_i^{mat}$ represents the force acting exclusively on the matter enclosed in the slab, whereas in a closed system $F_i^{tot}$ represents the time rate of change of the total momentum within the slab.

When the velocity of the moving bodies is enforced to be a constant through the application of an external force ($F_{i,x}^{ext} = -dp_{ps,i}/dt$) there is evidently a power flow into the system [see Eq. (3) with $\mathbf{j}_{ext} = 0$] and hence it is understandable that the fields may grow exponentially. Note that in presence of an exponentially growing oscillation, $\sum_i \mathbf{F}_{tot,i}^{ext} \cdot \mathbf{v}_i$ is also exponentially growing. Hence, $\sum_i \mathbf{F}_{tot,i}^{ext} \cdot \mathbf{v}_i$ must be positive because otherwise $H_{tot}$ would become negative for sufficiently large *t*, which contradicts $H_{tot} \geq 0$ (see Appendix A). Thus, for a growing oscillation there is a continuous power flow into the system. Therefore, despite $H_{EM,P}$ is time independent, the total energy (including the degrees of freedom associated with the translational motion of the system) increases with time. Moreover, even though the *wave energy* associated with a specific body can be negative (when **M** is indefinite) it turns out that the total energy stored in the body is always positive (see Eq. (A10) of Appendix A) [17].

## III. Friction force operator

In order to determine the friction force we expand the electromagnetic field in terms of natural modes of oscillation. It was proven in Ref. [17], under the assumption that the velocities are enforced to be constant, that the eigenmodes of the system are either associated with real-valued frequencies ($\mathbf{F}_{n\mathbf{k}} \leftrightarrow \omega_{n\mathbf{k}}$) or with complex-valued frequencies ($\mathbf{f}_{n\mathbf{k}} \leftrightarrow \omega_{c,n\mathbf{k}} = \omega'_{n\mathbf{k}} + i\lambda_{n\mathbf{k}}$ and $\mathbf{e}_{n\mathbf{k}} \leftrightarrow \omega^*_{c,n\mathbf{k}} = \omega'_{n\mathbf{k}} - i\lambda_{n\mathbf{k}}$). The eigenmodes $\mathbf{F}_{n\mathbf{k}}$, $\mathbf{f}_{n\mathbf{k}}$ and $\mathbf{e}_{n\mathbf{k}}$ are normalized to satisfy the orthogonality conditions:

$$\langle \mathbf{e}_{n\mathbf{k}} | \mathbf{e}_{m\mathbf{q}} \rangle = \langle \mathbf{f}_{n\mathbf{k}} | \mathbf{f}_{m\mathbf{q}} \rangle = \langle \mathbf{e}_{n\mathbf{k}} | \mathbf{F}_{m\mathbf{q}} \rangle = \langle \mathbf{f}_{n\mathbf{k}} | \mathbf{F}_{m\mathbf{q}} \rangle = 0, \tag{12a}$$

$$\langle \mathbf{e}_{n\mathbf{k}} | \mathbf{f}_{m\mathbf{q}} \rangle = \delta_{n,m} \delta_{\mathbf{k},\mathbf{q}}, \tag{12b}$$

$$\langle \mathbf{F}_{n\mathbf{k}} | \mathbf{F}_{m\mathbf{q}} \rangle = \pm \delta_{n,m} \delta_{\mathbf{k},\mathbf{q}}. \tag{12c}$$

where $\langle . | . \rangle$ denotes the indefinite inner product of Eq. (9). The electromagnetic modes $\mathbf{f}_{n\mathbf{k}}$ and $\mathbf{e}_{n\mathbf{k}}$ are related by $\mathbf{e}_{n\mathbf{k}} = \tilde{\mathbf{f}}_{n\mathbf{k}}$ with $\tilde{\mathbf{f}}$ defined by,

$$\tilde{\mathbf{f}}(\mathbf{r}) = \mathbf{U} \cdot \mathbf{f}^*(\mathbf{R}_{z,\pi} \cdot \mathbf{r}) \quad \text{with} \quad \mathbf{U} = \begin{pmatrix} \mathbf{R}_{z,\pi} & 0 \\ 0 & -\mathbf{R}_{z,\pi} \end{pmatrix}, \tag{13}$$

being $\mathbf{R}_{z,\pi} = -(\hat{\mathbf{x}}\hat{\mathbf{x}} + \hat{\mathbf{y}}\hat{\mathbf{y}}) + \hat{\mathbf{z}}\hat{\mathbf{z}}$ the transformation matrix associated with the 180º rotation around the z-axis. The electromagnetic field in the cavity can be expanded as $\hat{\mathbf{F}} = \begin{pmatrix} \hat{\mathbf{E}} & \hat{\mathbf{H}} \end{pmatrix}^T = \hat{\mathbf{F}}_R + \hat{\mathbf{F}}_C$ with:

$$\hat{\mathbf{F}}_R = \sum_{n\mathbf{k} \in E_R} \sqrt{\frac{\hbar |\omega_{n\mathbf{k}}|}{2}} \left( \hat{c}_{n\mathbf{k}} e^{-i\omega_{n\mathbf{k}} t} \mathbf{F}_{n\mathbf{k}} + \hat{c}^\dagger_{n\mathbf{k}} e^{+i\omega_{n\mathbf{k}} t} \mathbf{F}^*_{n\mathbf{k}} \right) \tag{14a}$$

$$\hat{\mathbf{F}}_C = \sum_{n\mathbf{k} \in E_C} \left( \hat{\beta}_{n\mathbf{k}} e^{-i\omega_{c,n\mathbf{k}} t} \mathbf{f}_{n\mathbf{k}} + \hat{\chi}_{n\mathbf{k}} e^{-i\omega^*_{c,n\mathbf{k}} t} \mathbf{e}_{n\mathbf{k}} + \hat{\beta}^\dagger_{n\mathbf{k}} e^{i\omega^*_{c,n\mathbf{k}} t} \mathbf{f}^*_{n\mathbf{k}} + \hat{\chi}^\dagger_{n\mathbf{k}} e^{i\omega_{c,n\mathbf{k}} t} \mathbf{e}^*_{n\mathbf{k}} \right) \tag{14b}$$

The coefficients of the expansion are $\hat{c}_{n\mathbf{k}}$, $\hat{\beta}_{n\mathbf{k}}$, and $\hat{\chi}_{n\mathbf{k}}$. The hat "^" indicates that in the framework of a quantum theory the pertinent symbol should be understood as an operator. In the framework of a classical theory, the coefficients $\hat{c}_{n\mathbf{k}}$, $\hat{\beta}_{n\mathbf{k}}$, and $\hat{\chi}_{n\mathbf{k}}$ are scalars. The summations associated with the real-valued (complex-valued) eigenvalues are restricted to the sets $E_R$ ($E_C$) such that $E_R = \{(n,\mathbf{k}): \omega_{n\mathbf{k}} \langle \mathbf{F}_{n\mathbf{k}} | \mathbf{F}_{n\mathbf{k}} \rangle > 0\}$ and $E_C = \{(n,\mathbf{k}): \lambda_{n\mathbf{k}} = \text{Im}\{\omega_{c,n\mathbf{k}}\} > 0 \text{ and } k_x > 0\}$. In Appendix B, we calculate the contribution of a generic term of the series (14) to the friction force. As could be anticipated, oscillators associated with a real valued frequencies (in the set $E_R$) do not contribute to the force. On the other hand, a pair of oscillators associated with the complex-valued frequencies $\omega_c$ and $\omega_c^*$ exerts a friction force over the $i$-th body given by (for simplicity the index $n\mathbf{k}$ is omitted below) [see Eq. (B8)]:

$$\frac{d\hat{p}_{u,i}}{dt} = 2\lambda p_{u,i}(\mathbf{f})\left[e^{2\lambda t}\left(\hat{\beta}\hat{\beta}^\dagger + \hat{\beta}^\dagger\hat{\beta}\right) - e^{-2\lambda t}\left(\hat{\chi}\hat{\chi}^\dagger + \hat{\chi}^\dagger\hat{\chi}\right)\right], \quad u=wm, ps \quad (15)$$

where $p_{wv,i}(\mathbf{f}) = \frac{1}{2}\int_{V_i}\left(\mathbf{D_f}\times\mathbf{B_f^*} + \mathbf{D_f^*}\times\mathbf{B_f}\right)\cdot\hat{\mathbf{x}}d^3\mathbf{r}$ is the wave momentum associated with the complex-valued field, $p_{EM,i}(\mathbf{f}) = \frac{1}{2c^2}\int_{V_i}\left(\mathbf{E_f}\times\mathbf{H_f^*} + \mathbf{E_f^*}\times\mathbf{H_f}\right)\cdot\hat{\mathbf{x}}d^3\mathbf{r}$ is the electromagnetic momentum, and $p_{ps,i}(\mathbf{f}) = p_{wm,i}(\mathbf{f}) - p_{EM,i}(\mathbf{f})$. We recall that $d\hat{p}_{ps,i}/dt$ is the friction force acting on the matter enclosed in the $i$-th slab, whereas in a closed system $d\hat{p}_{wv,i}/dt$ is the time rate of change of the total momentum within the slab. Hence, provided $p_{wv,i}(\mathbf{f}) \neq 0$ ($p_{ps,i}(\mathbf{f}) \neq 0$) it is evident that $F_i^{tot}$ ($F_i^{mat}$) does not vanish in presence of system instabilities associated with complex-valued frequencies of oscillation. It should be emphasized that this conclusion is valid for both classical and quantum systems, and thus the friction force has a classical counterpart.

In quantum theory the field amplitudes $\hat{\beta}$, and $\hat{\chi}$ associated with a pair of complex-valued frequencies are written in terms of annihilation and creation operators $\hat{a}_c, \hat{b}_c$ and $\hat{a}_c^\dagger, \hat{b}_c^\dagger$, satisfying the standard commutation relations (shown below) [17]:

$$\hat{\beta} = \frac{1}{2}\sqrt{\hbar\omega_c}\left(\hat{a}_c + \hat{b}_c^\dagger\right), \qquad \hat{\chi} = \frac{1}{2}\sqrt{\hbar\omega_c^*}\left(\hat{a}_c - \hat{b}_c^\dagger\right) \qquad (16a)$$

$$\left[\hat{a}_c, \hat{a}_c^\dagger\right] = \left[\hat{b}_c, \hat{b}_c^\dagger\right] = 1, \qquad \left[\hat{a}_c, \hat{b}_c\right] = \left[\hat{a}_c, \hat{b}_c^\dagger\right] = 0 \qquad (16b)$$

Substituting this result in Eq. (15) and summing over all the complex-valued oscillators (crossed terms associated with possible contributions from oscillators $n\mathbf{k}$ and $m\mathbf{k}$, with $n \neq m$, are neglected; the quantum expectation of $\hat{F}_i^{tot}$ is independent of the crossed terms) it is found that (see also Eq. (B10)):

$$\hat{F}_i^{tot} = \sum_{\lambda_{n\mathbf{k}}>0 \text{ and } k_x>0} 2\hbar\lambda_{n\mathbf{k}}|\omega_{c,n\mathbf{k}}|p_{wv,i}(\mathbf{f}_{n\mathbf{k}}) \times \\ \left[\left(\hat{a}_{c,n\mathbf{k}}\hat{a}_{c,n\mathbf{k}}^\dagger + \hat{b}_{c,n\mathbf{k}}^\dagger\hat{b}_{c,n\mathbf{k}}\right)\sinh 2\lambda_{n\mathbf{k}}t + \left(\hat{a}_{c,n\mathbf{k}}\hat{b}_{c,n\mathbf{k}} + \hat{b}_{c,n\mathbf{k}}^\dagger\hat{a}_{c,n\mathbf{k}}^\dagger\right)\cosh 2\lambda_{n\mathbf{k}}t\right]. \qquad (17)$$

The friction force acting on the matter enclosed by the $i$-th slab is given by a similar formula with $p_{wv,i}(\mathbf{f}_{n\mathbf{k}})$ replaced by $p_{ps,i}(\mathbf{f}_{n\mathbf{k}})$. It is important to stress that this result is derived under the hypothesis that the velocity of the moving slabs is kept constant in the time window of interest, which can be ensured either by applying an external force ($F_{i,x}^{ext} = -dp_{ps,i}/dt$), or by considering very massive bodies such that a change in the kinetic momentum results in a insignificant change of the velocity.

## IV. Selection rules

As demonstrated in the previous section, the friction force is a consequence of system instabilities manifested in the form of natural modes of oscillation with complex valued frequencies. In our previous work [17], we have shown using perturbation theory that these modes are the result of the interaction of guided waves supported by the moving bodies. In Appendix C, considering the limit of a weak interaction and that $v_i/c \ll 1$, we further develop these ideas and obtain "selection rules" for the interacting guided modes that give rise to oscillations with complex valued frequencies. The picture that emerges from our model reveals that the system instabilities are the result of the hybridization of two guided waves (each attached to a given body) such that *(i)* the interacting guided waves are associated with the same frequency of oscillation $\omega$ and the same wave vector $(k_x, k_y)$ in a fixed reference frame. *(ii)* the frequencies of oscillation of the guided modes, when calculated in the frame co-moving with the relevant dielectric body, are such that they have opposite signs.

Specifically, let $\omega_1 = \tilde{\omega}_1 + v_1 k_x$ and $\omega_2 = \tilde{\omega}_2 + v_2 k_x$ be the Doppler shifted frequencies (as measured in the lab frame), where $\tilde{\omega}_i = \tilde{\omega}_i(k_x, k_y)$ represents the dispersion of a guided mode in the frame co-moving with the *i-th* slab. In the non-relativistic limit $k_x$ stays invariant under a change of the reference frame. It is demonstrated in Appendix C [Eqs. (C4) and (C5)] that the hybridization of the two guided modes results in a natural mode with a complex valued frequency provided the following selection rules are satisfied:

$$\tilde{\omega}_1 + v_1 k_x = \tilde{\omega}_2 + v_2 k_x. \tag{18a}$$

$$\tilde{\omega}_1 \tilde{\omega}_2 < 0 \tag{18b}$$

Note that it is implicit that $(k_x, k_y)$ is the same for the two interacting guided modes. Let us denote the "phase" refractive index associated with the guided mode of the *i-th* slab in the respective co-moving frame as $n_{ph,i} = ck/\tilde{\omega}_i$, where $k = \sqrt{k_x^2 + k_y^2}$. Then, the selection rule (18a) implies that:

$$\frac{1}{c}(v_2 - v_1) = -\frac{\sqrt{k_x^2 + k_y^2}}{k_x}\left(\frac{1}{n_{ph,2}} - \frac{1}{n_{ph,1}}\right). \tag{19}$$

Note that $n_{ph,i}$ may be either positive or negative depending on the sign of $\tilde{\omega}_i$, and because of Eq. (18b) $n_{ph,1} n_{ph,2} < 0$. For waveguides based on non-dispersive materials (e.g. a grounded dielectric slab), $n_{ph,i}$ is always smaller than the refractive index of the dielectric: $|n_{ph,i}| < n_i$. Thus, the minimum of the absolute value of the right-hand side of Eq. (19) is attained for $k_y = 0$ and $n_{ph,i} = \pm n_i$. Hence, to have a friction force the velocities of the two moving slabs must satisfy:

$$|v_2 - v_1| \geq c \left( \frac{1}{n_2} + \frac{1}{n_1} \right) \qquad \text{(threshold for friction).} \qquad (20)$$

Thus, at least one of the bodies moves with a velocity larger than the corresponding Cherenkov threshold in the laboratory frame.

To further develop these ideas, we consider the particular case wherein the dielectric slabs are identical. For each slab, the guided modes are characterized in co-moving frame by the dispersion branches $+\tilde{\omega}^{(n)}$ (positive frequencies) and $-\tilde{\omega}^{(n)}$ (negative frequencies). The index $n$ labels the guided modes branches. The simplest possibility (but not the only one) to satisfy the selection rules (18) is to consider the interaction of modes associated with the same $n$ such that $\tilde{\omega}_1 = \pm \tilde{\omega}^{(n)}$ and $\tilde{\omega}_2 = \mp \tilde{\omega}^{(n)}$. In that case Eq. (19) implies that

$$n_{ph} = \pm \frac{\sqrt{k_x^2 + k_y^2}}{k_x} \frac{2c}{v_2 - v_1}$$ being $n_{ph} = ck / \tilde{\omega}^{(n)}$. Noting that $n_{ph}$ depends on $(k_x, k_y)$ as $n_{ph} = n_{ph}\left(\sqrt{k_x^2 + k_y^2}\right)$ it can be verified that the previous equation has always a solution when $n_{ph} \geq 2c/|v_2 - v_1|$. This demonstrates that when $n_{ph} \geq 2c/|v_2 - v_1|$ there is always a suitable $(k_x, k_y)$ that satisfies the selection rules.

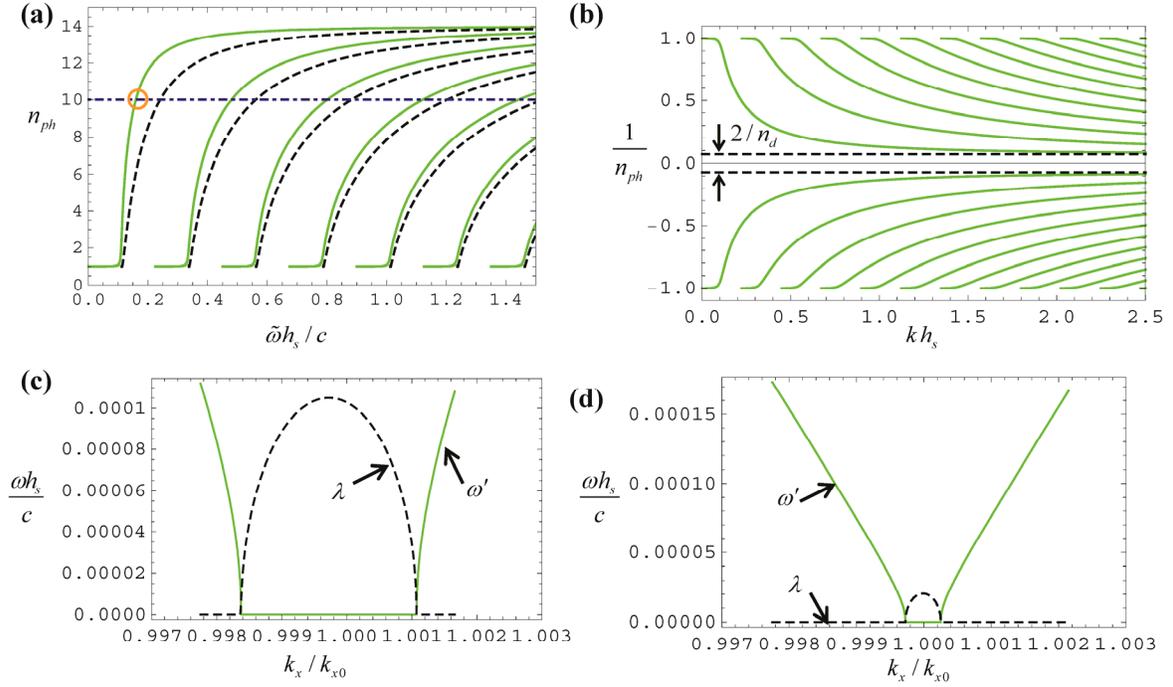

Fig. 2. (Color online) (a) Dispersion diagram ($n_{ph} = ck/\tilde{\omega}$ vs. $\tilde{\omega}$) for the *p*-polarized guided modes (solid green curves) and *s*-polarized guided modes (dashed black curves) supported by a grounded dielectric slab with thickness $h_s$ and refractive index $n_d = 14$. The frequency $\tilde{\omega}$ is measured in the co-moving frame. The guided modes lying above the dot-dashed blue horizontal line ($n_{ph} = 10$) in the dispersion diagram are the ones responsible for the friction force when two identical slabs are in relative motion with $v_2 = -v_1 = c/10$. (b) Similar to (a) for *p*-polarized waves, but $1/n_{ph}$ is plotted as a function of $k$. (c) Dispersion of the natural mode ($\omega = \omega' + i\lambda$ vs. $k_x$ with $k_y = 0$) resulting from the hybridization of two *p*-polarized guided modes of the individual slabs with $n_{ph} = \pm 10$ (the pertinent mode with $n_{ph} = 10$ is marked in panel *a* with an orange circle; $k_{x0} = 1.598/h_s$ is the wave number associated with this mode) for $v_2 = -v_1 = c/10$ and $d = h_s$. (d) Similar to (c) but for $d = 2h_s$.

To illustrate the discussion, we show in Fig. 2a the dispersion of the guided modes supported by a single (non-magnetic) dielectric slab with thickness $h_s$ with refractive index $n_d = 14$ surrounded by a vacuum. The dielectric slab is backed by a PEC ground

-17-

plane, similar to what is illustrated in Fig. 1a. Consistent with the previous discussion, it is seen that $1 < |n_{ph}| < n_d$ and that the guided modes have several dispersion branches. In Fig. 2b we represent $1/n_{ph}$ as a function of $k$, showing also the negative frequency branches. If the vertical distance between two branches associated with the same $k$ and having $n_{ph}$ with different signs is smaller than $(v_2 - v_1)/c$ then the hybridization of the corresponding modes may result in a system instability.

This property is demonstrated in Figs. 2c and 2d which depict the dispersion of the $\lambda > 0$ mode associated the hybridization of two *p*-polarized guided modes of the individual slabs with $n_{ph} = \pm 10$ (the relevant mode with $n_{ph} = 10$ is marked in Fig. 2a with an orange circle) for the case $v_2 = -v_1 = c/10$ and two values of the slabs distance *d*. The dispersion of the hybridized guided mode is computed using Eq. (C1) of Appendix C. The plots show that when the selection rules are satisfied (this corresponds to the condition $k_x / k_{x0} = 1$) the hybridized modes have $\omega = \omega' + i\lambda$ such that $\lambda \neq 0$. The value of $\lambda$ is larger for smaller *d*, i.e. in a case of a stronger modal interaction. When $k_x / k_{x0}$ is slightly different from the unity the selection rules are not satisfied and thus the hybridized modes have $\lambda = 0$. Note that in the laboratory frame $\omega' \approx 0$ because of the Doppler effect, i.e. $\tilde{\omega} + v_1 k_x = \omega_1 \approx \omega_2 = -\tilde{\omega} + v_2 k_x \approx 0$. This is a general consequence of the selection rule (18a) when $\tilde{\omega}_1 = -\tilde{\omega}_2$ and $v_1 = -v_2$. We verified (not shown here) that if the effect of loss is considered in the material response it is still possible to have natural modes associated with growing oscillations (see also Ref. [18]). This result is also evident because of analytic continuation arguments. We also verified that relativistic corrections result in a small shift of the modal diagrams.

For future reference, it is mentioned that the sign of the pseudo-momentum associated with a system instability is such that in the *i-th* slab

$$\text{sgn}(p_{ps,i}) = -\text{sgn}(v_i - v_j), \qquad \text{where } i,j = 1,2 \text{ and } i \neq j. \qquad (21)$$

The proof is given in Appendix D.

## V. Quantum expectation of the friction force

In the framework of a quantum theory, a generic state of the system is a superposition of states of the type $\left|...,\left(m^a_{c,n\mathbf{k}}, m^b_{c,n\mathbf{k}}\right),...,m^c_{n\mathbf{k}},...\right\rangle$, where $\left(m^a_{c,n\mathbf{k}}, m^b_{c,n\mathbf{k}}\right)$ are the occupation numbers of the oscillators associated with $\hat{a}_{c,n\mathbf{k}}, \hat{b}_{c,n\mathbf{k}}$, and $m^c_{n\mathbf{k}}$ is the occupation number associated with the oscillators associated with real-valued frequencies ($\hat{c}_{n\mathbf{k}}$) [17]. When the "wave part" of the system is electromagnetically unstable it has a peculiar property: there are no stationary states, and in particular there is no ground state [17]. However, it was shown in Ref. [17] that – provided $\mathbf{e}_{n\mathbf{k}}$ is chosen as in Eq. (13) – the state wherein all the occupation numbers (for oscillators associated with either real-valued or complex valued frequencies) vanish, $|\Omega\rangle = \left|...,\left(0^a_{c,n\mathbf{k}}, 0^b_{c,n\mathbf{k}}\right),...\right\rangle$ is such that the oscillations of the quantized fields are minimal. Thus, $|\Omega\rangle$ is a *pseudo-ground* because it corresponds to the state of minimal disturbance of the "wave part" of the system (in the sense that the oscillation amplitudes are minimal), despite it is not the state of minimal wave energy. Using the formula for $\hat{F}^{tot}_i$ [Eq. (17)] we can determine the expectation of the operator ($\langle \hat{F}^{tot}_i \rangle$) supposing that the system is prepared in the pseudo-ground $|\Omega\rangle$ at *t*=0. We find that

$$\left\langle \hat{F}_i^{tot} \right\rangle = \sum_{\lambda_{n\mathbf{k}}>0 \text{ and } k_x>0} 2\hbar \lambda_{n\mathbf{k}} \left| \omega_{c,n\mathbf{k}} \right| p_{wv,i}\left( \mathbf{f}_{n\mathbf{k}} \right) \sinh 2\lambda_{n\mathbf{k}} t, \qquad t \geq 0. \qquad (22)$$

It is proven in Appendix D that since the moving bodies have a large refractive index ($n_i \gg 1$) the force acting on the matter enclosed by the $i$-th slab is only marginally smaller than the rate of change of the total momentum in $i$-th slab, so that $\left\langle \hat{F}_i^{mat} \right\rangle \approx \left\langle \hat{F}_i^{tot} \right\rangle$. The correction factor is of the order of $1 - 1/n_i^2$ [Eq. (D6)] at the friction force threshold. Thus, we can identify $\left\langle \hat{F}_i^{tot} \right\rangle$ with the friction force, and this is done in what follows.

Notably, it is seen that at $t=0$ the expectation of the friction force vanishes: $\left\langle \hat{F}_i^{tot} \right\rangle_{t=0} = 0$. This is consistent with our picture of the pseudo-ground as the state of minimal interaction of the moving bodies. However, the pseudo-ground is analogous to a point of unstable equilibrium: as time passes the system is pushed away from the pseudo-ground and the friction forces builds up exponentially due to the generation of electromagnetic quanta (i.e. in the Schrödinger picture the occupation numbers associated with the operators $\hat{a}_{c,n\mathbf{k}}, \hat{b}_{c,n\mathbf{k}}$ increase with time [17]). This picture of the friction force provided by our quantum theory (illustrated in Fig. 1b) differs markedly from the semi-classical theory of Pendry and from other approaches based on first order perturbation theory, wherein the friction force is constant and independent of time [1, 4, 8]. It is emphasized that the exponential growth of the force can only be sustained provided the velocity of the moving bodies is kept constant, which as discussed previously requires an external action. In general, the friction force provides the feedback mechanism that prevents the indefinite growth of the oscillations of the electromagnetic field.

It is possible to reconcile our quantum theory with the results of Pendry. To do this, we start by noting that in Pendry's approach [1, 4] the friction force is computed based on Fermi's golden rule and on the number of excitations generated per unit of time. It was shown in our previous article [17] that supposing the initial state is the pseudo-ground $\left|0^a_{c,n\mathbf{k}}, 0^b_{c,n\mathbf{k}}\right\rangle$ the probability of the system being in the state $\left|0^a_{c,n\mathbf{k}}, 0^b_{c,n\mathbf{k}}\right\rangle$ at a later time instant is exactly $\text{sech}^2(\lambda_{n\mathbf{k}} t)$ [17, Eq. (66)]; here, we consider a single pair of coupled oscillators associated with $\hat{a}_{c,n\mathbf{k}}, \hat{b}_{c,n\mathbf{k}}$. This shows that the average time to generate an excitation is of the order $t \sim 1/(2\lambda_{n\mathbf{k}})$, which can be regarded as the "lifetime" of the initial state. Thus, the rate of excitations for the considered pair of oscillators is equal to $R_{g,n\mathbf{k}} = 2\lambda_{n\mathbf{k}}$. This rate should be comparable with what is obtained with Fermi's golden rule in Pendry's approach. Now, one crucial point is that in Pendry's theory it is implicit that the lifetime of the generated excitations is very short (e.g. they are quickly absorbed due to loss in the system) as compared to $1/(2\lambda_{n\mathbf{k}})$, because the system is assumed to be always in the ground state. Quite differently, in our approach the dielectrics are lossless and hence the generated excitations stay in the system and promote new excitations making $R_{g,n\mathbf{k}}$ effectively time dependent. This is why in our theory the force grows exponentially, whereas in Pendry's theory the force is constant.

Since the average time to generate an excitation is of the order of $1/(2\lambda_{n\mathbf{k}})$, we see that Pendry's force (for each pair of oscillators) should be identified with the force calculated with our theory at the time instant $t \approx 1/(2\lambda_{n\mathbf{k}})$. Hence, supposing for simplicity that the ground states of all the relevant oscillators have similar lifetimes, and using $\sinh 1 = 1.18 \approx 1$ we get the following formula for the friction force:

$$\left\langle \hat{F}_i^{tot} \right\rangle_{1,g} \approx \sum_{\lambda_{n\mathbf{k}} > 0 \text{ and } k_x > 0} 2\hbar \lambda_{n\mathbf{k}} \left| \omega_{c,n\mathbf{k}} \right| p_{wv,i}(\mathbf{f}_{n\mathbf{k}}) \tag{23}$$

The force $\left\langle \hat{F}_i^{tot} \right\rangle_{1,g}$ can be regarded as the force calculated at the time instant corresponding to the creation of the first excitation. Using, Eqs. (21) and (D5) it is also possible to write that:

$$\left\langle \hat{F}_i^{tot} \right\rangle_{1,g} \approx -\mathrm{sgn}(v_i - v_j) \sum_{\lambda_{n\mathbf{k}} > 0 \text{ and } k_x > 0} 2\hbar \lambda_{n\mathbf{k}} \left| \omega_{c,n\mathbf{k}} \right| \left| p_{wv,i}(\mathbf{f}_{n\mathbf{k}}) \right| \tag{24}$$

where $i, j = 1, 2$ and $i \neq j$. This confirms that the friction force acts against the relative motion of the two slabs. In Appendix C, we give a detailed proof that $\left\langle \hat{F}_i^{tot} \right\rangle_{1,g}$ exactly coincides with the calculation of Pendry in the limit of a weak interaction between the moving bodies [1, 4]. This validates the previous discussion, and demonstrates that the semiclassical theory of Pendry can be recovered from our dynamic quantum theory. It is interesting to mention that the absolute value of $\left\langle \hat{F}_i^{tot} \right\rangle_{1,g}$ is exactly coincident with the variance of the force, $\left| \hat{F}_i^{tot} \right| = \left\langle \left( \hat{F}_i^{tot} \right)^2 \right\rangle_{t=0}^{1/2}$, calculated at $t = 0$. Indeed, straightforward calculations show that:

$$\left| \hat{F}_i^{tot} \right| = \sum_{\lambda_{n\mathbf{k}} > 0 \text{ and } k_x > 0} 2\hbar \lambda_{n\mathbf{k}} \left| \omega_{c,n\mathbf{k}} \right| \left| p_{wv,i}(\mathbf{f}_{n\mathbf{k}}) \right| = \left| \left\langle \hat{F}_i^{tot} \right\rangle_{1,g} \right|. \tag{25}$$

To numerically evaluate $\left| \hat{F}_i^{tot} \right|$ we need to know all natural modes associated with complex valued frequencies ($\mathbf{f}_{n\mathbf{k}} \leftrightarrow \omega_{c,n\mathbf{k}} = \omega'_{n\mathbf{k}} + i\lambda_{n\mathbf{k}}$). It is proven in Appendix D [Eq. (D1)] that for the geometry of Fig. 1a and weakly interacting slabs $\left| p_{wv,i} \left( \mathbf{f}_{n\mathbf{k}} \right) \right| \approx \left| k_x \right| / \left| 2\omega_{c,n\mathbf{k}} \right|$. Hence, in this limit, and assuming also that the slabs are infinitely wide so that $(k_x, k_y)$ varies in a continuum ( $\sum_{(k_x,k_y)} \to \frac{A_0}{(2\pi)^2} \iint dk_x dk_y$, being $A_0$ the transverse area of the slabs in the *xoy* plane), we find that:

$$\frac{\left| \hat{F}_i^{tot} \right|}{A_0} = \frac{\hbar}{(2\pi)^2} \int_{-\infty}^{+\infty} dk_y \int_0^{+\infty} dk_x\, k_x \left( \sum_{\lambda_{n\mathbf{k}}>0} \lambda_{n\mathbf{k}} \right). \tag{26}$$

Thus, the friction force per unit of area depends mainly on the imaginary part of the frequencies of oscillation. It is possible to evaluate the sum in inner brackets in the previous formula with the help of the argument principle of complex analysis. Specifically, let $D(\omega, \mathbf{k})$ be such that $D(\omega, k_x, k_y) = 0$ is the characteristic equation for the natural modes of the system. A formal expression for $D$ is given in Appendix C for the problem of two interacting moving bodies backed with a perfectly electrical conducting layer. The argument principle – also widely used in calculations of the Casimir force [26-29] – implies that:

$$\sum_{\lambda_{n\mathbf{k}}>0} \lambda_{n\mathbf{k}} = \mathrm{Im}\left\{ \frac{1}{2\pi i} \oint_C \frac{D'(\omega,\mathbf{k})}{D(\omega,\mathbf{k})} \omega\, d\omega \right\} \tag{27}$$

where $D' = \partial_\omega D$ and $C$ is a contour that encircles the upper half plane (UHP), passing right above the real positive axis and that joins $\omega = +\infty$ and $\omega = -\infty$ with a semicircle of infinite radius. The integral over the semicircle can be discarded because it does not contribute to the friction force. Hence, substituting Eq. (27) into Eq. (26), we find that:

$$\frac{\left| \hat{F}_i^{tot} \right|}{A_0} = \frac{-\hbar}{(2\pi)^3} \mathrm{Re}\left\{ \int_{-\infty}^{+\infty} dk_y \int_0^{+\infty} dk_x \int_{-\infty}^{+\infty} d\omega\, k_x \omega \frac{D'(\omega,\mathbf{k})}{D(\omega,\mathbf{k})} \right\}. \tag{28}$$

It is relevant to mention that even though Eq. (28) was derived under the assumption that the considered dielectric bodies are made of dispersionless dielectrics, Eqs. (25)-(28) can be readily extended to the general case wherein the considered bodies are made of dispersive lossy dielectrics. This suggests that our theory may apply in a broader context.

## VI. Numerical Examples

To illustrate the application of the theory, we computed $\left|\hat{F}_i^{tot}\right|$ for the geometry of Fig. 1a supposing in all the examples that $v_2 = -v_1$, and that the two slabs are identical. The slabs are made of a nonmagnetic material $(\mu = \mu_0)$ with refractive index $n_d$, and have the thickness $h_s \equiv h_1 = h_2$. To have further confidence in the numerical calculations, we computed the force based on two different formulas *(i)* Eq. (28) calculating the dispersion $D(\omega, \mathbf{k})$ of the hybridized modes with Eq. (C1) of Appendix C. *(ii)* Eq. (C19) derived in Appendix C, wherein the friction force is directly written in terms of the dispersion of the guided modes supported by the individual slabs in the respective co-moving frames. This latter approach is less rigorous for small values of *d* and only takes into account the contributions of guided modes with $\tilde{\omega}_1 = -\tilde{\omega}_2$. In the non-relativistic limit the hybridized modes split into *p*-polarized and *s*-polarized waves (see Appendix C), and thus the total force can be seen as the sum of two independent components, $\left|F^{tot}\right| = \left|F^{tot,p}\right| + \left|F^{tot,s}\right|$, being $\left|F^{tot,p}\right|$ ($\left|F^{tot,s}\right|$) the contribution to the force of *p* (*s*) polarized hybridized modes.

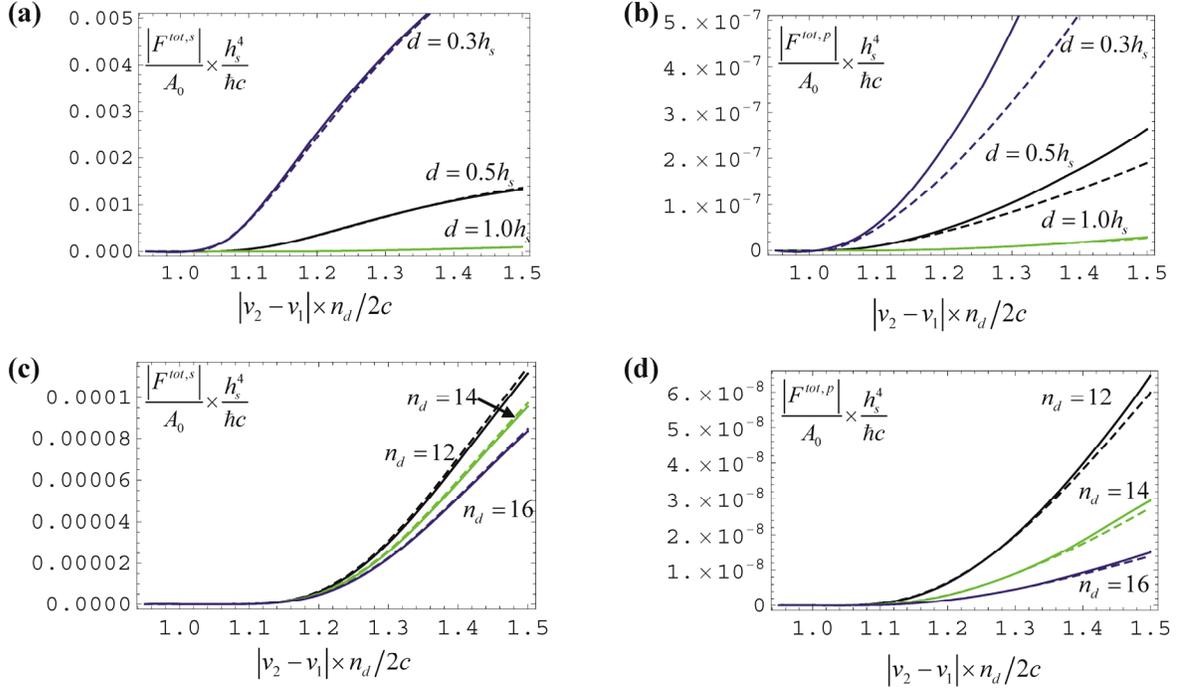

Fig. 3. (Color online) Normalized friction force as a function of the normalized $|v_2 - v_1|$ for two identical dielectric moving bodies with velocities $v_2 = -v_1$. (a) Component of the force due to s-polarized waves for several slab distances $d$. The refracting index of the dielectrics in the respective co-moving frames is $n_d = 14$. (b) Similar to (a) but for p-polarized waves. (c) Similar to (a) but for the case $d = h_s$ and for several refractive indices $n_d$. (d) Similar to (c) but for p-polarized waves. In the plots the solid lines refer to the calculation obtained with the exact formula (28), whereas the dashed lines where obtained with the simplified formula (C19).

In Fig. 3 we depict the calculated normalized friction force as a function of the normalized velocity $|v_2 - v_1|$ for different values of the distance $d$ and of the refractive index $n_d$. A general observation is that the two formulas used for the computation of the friction force give very consistent results, particularly when $d/h_s$ has large values. As expected, the friction force decreases significantly when the normalized distance between the moving bodies increases. Also, consistent with Eq. (20), it is seen that the friction force is non-zero only for normalized velocities such that $|v_2 - v_1|n_d/2c > 1$. Figures 3c and 3d also reveal that for fixed $|v_2 - v_1|n_d/2c$ the force decreases with increasing $n_d$. However, one should keep in mind that for a fixed $|v_2 - v_1|n_d/2c$ slabs with smaller $n_d$ move faster. Typically, for a fixed $|v_2 - v_1|/2c$ the force grows with increasing $n_d$.

It is striking from Fig. 3 that $|F^{tot,s}|$ can be several orders of magnitude larger than $|F^{tot,p}|$. This is to some extent surprising because the *p*-polarized modes have a dispersion branch with no frequency cut-off, unlike the *s*-polarized modes which can only propagate above a certain threshold frequency (see Fig. 2a). A consequence of this is that for a fixed $|v_2 - v_1|/2c$ the *p*-polarized modes that contribute to the force have smaller $k = \sqrt{k_x^2 + k_y^2}$, than the *s*-polarized modes that contribute to the force. Because the interaction between the slabs is mediated by a term of the type $e^{-2kd}$ [see Eq. (C1)], one might expect that the friction force would be mainly determined by *p*-polarized waves, in contradiction with the numerical results. However, it turns out that the hybridization of the guided modes is much stronger for *s*-polarized waves. To explain why in simple physical terms, we consider that the relevant guided modes propagate along *x*. Then, the strength of the hybridization is mainly determined by the interactions of the field $H_y$ of slab 1 and 2 in case of *p*-polarized waves, and of $E_y$ in case of *s*-polarized waves. The key point is that because $n_d \gg 1$ the dielectric-vacuum interface is effectively seen as a magnetic wall from the dielectric side, and thus $H_y$ is quite small at the dielectric-vacuum interface whereas typically $E_y$ has a maximum at the same interface. As a consequence, the hybridization of *s*-polarized modes is more efficient and leads to a larger $\lambda = \text{Im}\{\omega_c\}$, and thus to a stronger force. Note that from the stress tensor theorem [21] the friction force should be related to the value of $\varepsilon_0 E_z E_x$ for *p*-polarized waves, and $\mu_0 H_z H_x$ for *s*-polarized waves, calculated at the vacuum side of the interfaces, and thus our results indicate that, in the considered example, the friction force has a magnetic origin.

To have an idea of the magnitude of the quantum friction force, we consider that $n_d = 14$, $d = h_s = 1 \mu m$ and $|v_2 - v_1| n_d / 2c = 1.4$. In such a case, despite the large velocity of the moving bodies, the force per unit of area due to the interaction of s-polarized waves is only $|F^{tot,s}| / A_0 = 1.9 \times 10^{-6} N/m^2$. For $d/h_s$ fixed the force scales with $1/h_s^4$.

As mentioned at the end of Sect. V, there is no difficulty in applying formula (28) to other physical systems, even if the moving slabs are made of dispersive lossy dielectrics. To illustrate this, we consider the interaction of two moving metal sheets with $v_2 = -v_1$ and characterized by the reflection coefficient $R(\omega) = \dfrac{-\omega_{sp}^2}{\omega^2 - \omega_{sp}^2}$, being $\omega_{sp}$ the frequency associated with the surface plasmon resonance. For simplicity, in this example the dependence of *R* on the wave vector $(k_x, k_y)$ is ignored, similar to what was done in Ref. [4]. In Figure 4 we plot the friction force variance computed using the theory of this work [Eq. (28) using $D(\omega, \mathbf{k}) = 1 - e^{-2kd} R(\omega - k_x v_1) R(\omega - k_x v_2)$], and compare the result with the force computed based on the theory of Pendry [Eq. (15) of Ref. [4] multiplied by ¼; the reason for this multiplicative factor is that apparently Eq. (15) of Ref. [4] was derived assuming $R(\omega) = \dfrac{-2\omega_{sp}^2}{\omega^2 - \omega_{sp}^2}$, as Eq. (14) of the same article indicates]. As seen, the two approaches concur rather well, particularly for larger values of $d$ when the interaction between the moving metal sheets is weaker.

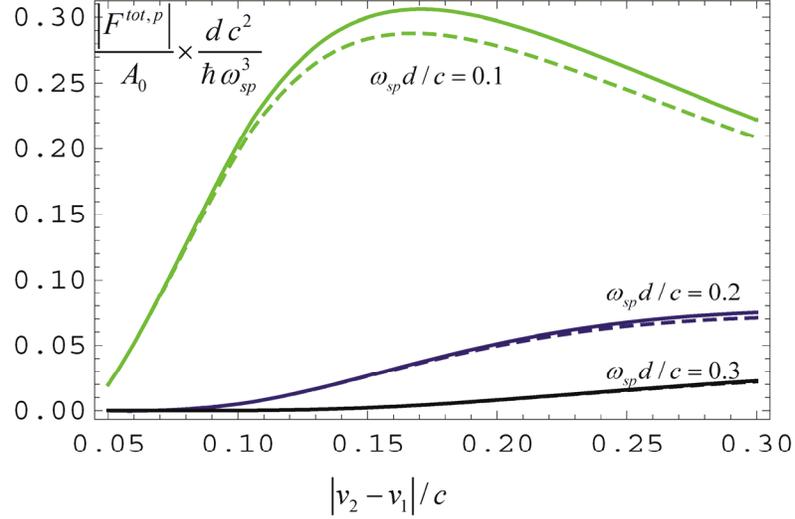

Fig. 4. (Color online) Normalized friction force as a function of $|v_2 - v_1|$ for two identical moving metal sheets with velocities $v_2 = -v_1$, for different normalized values of the distance $d$ between the bodies. Solid lines: Results obtained with the theory of this work [Eq. (28)]. Dashed lines: Results obtained with the theory of Pendry [4].

## VII. Conclusion

A first principles derivation of the quantum friction force was presented. It was proven that the expectation of the quantum friction force vanishes at the initial time instant for a system of non-dispersive moving bodies prepared in the "pseudo-ground" state. However, as time passes the expectation of the quantum friction force exponentially grows, as long as the change in the velocity of the moving bodies is insignificant. We calculated the quantum friction force at the time instant corresponding to the generation of the first excitation for each pair of unstable oscillators, and demonstrated that in the limit of a weak interaction it is coincident with the semiclassical result of Pendry [1, 4]. The velocity threshold above which quantum friction can take place was derived, and the effect of quantum friction was linked to system instabilities that may occur when at least one of the dielectric bodies has velocity larger than the Cherenkov emission threshold. These instabilities are due to the hybridization of guided modes supported by the individual bodies, and take place when the guided modes satisfy certain selection rules. We numerically estimated the quantum friction when two non-magnetic grounded dielectric slabs are in relative motion, and demonstrated that, surprisingly, in the non-relativistic limit the quantum friction is mainly determined by $s$-polarized waves. It is relevant to mention that the friction mechanism described in this article cannot be used to extract an infinite amount of energy from the system or achieve a "perpetuum mobile" [10]. Indeed, the "perpetuum mobile" argument of Ref. [10] relies on the hypothesis that we can have an ordinary passive magneto-electric material with the same constitutive relations as those of a moving slab. Clearly, in presence of wave instabilities this is impossible unless the magneto-electric material is active.

**Acknowledgements:** This work is supported in part by Fundação para a Ciência e a Tecnologia grant number PTDC/EEI-TEL/2764/2012.

## Appendix A:

In this Appendix, we study the electrodynamics of moving *rigid* bodies in the non-relativistic limit. The analysis is in part related to Ref. [30], which investigates the same problem but for moving electric dipoles, while here we consider the continuous limit.

As a starting point, we note that the electromagnetic field dynamics is determined by:

$$\nabla \times \mathbf{E} = -\partial_t \mathbf{B}, \qquad \nabla \times \mathbf{H} = \partial_t \mathbf{D} + \mathbf{j}_{ext}, \qquad \nabla \cdot \mathbf{D} = \rho_{ext}, \qquad (A1a)$$

where $\mathbf{j}_{ext}$ is an hypothetical external electric current density, and $\rho_{ext}$ is the corresponding external electric charge density, with $\nabla \cdot \mathbf{j}_{ext} + \partial_t \rho_{ext} = 0$. We consider a set of rigid material bodies that only have an electric response. Let $\mathbf{u}_i(t)$ be the coordinates of the center of mass of the $i$-th body, and $\mathbf{v}_i = d\mathbf{u}_i/dt$ be the center of mass velocity. For simplicity, we neglect possible torques, and thus only translational motions are allowed.

Let the electric polarization vector of the $i$-th body in the respective co-moving frame be denoted by $\mathbf{P}_{e,i}^{co}(\mathbf{R},t)$. Here, $\mathbf{R} = \mathbf{r} - \mathbf{u}_i$ are the local coordinates in the co-moving frame. It is understood that $\mathbf{P}_{e,i}^{co}(\mathbf{R},t)$ vanishes outside the volume of the $i$-th body. Then, within a non-relativistic approximation, the electric displacement and the magnetic field in the laboratory frame are given by:

$$\mathbf{D}(\mathbf{r},t) = \varepsilon_0 \mathbf{E}(\mathbf{r},t) + \mathbf{P}_e(\mathbf{r},t), \qquad \mathbf{H}(\mathbf{r},t) = \mu_0^{-1} \mathbf{B}(\mathbf{r},t) - \mu_0^{-1} \mathbf{P}_m(\mathbf{r},t). \qquad (A1b)$$

where $\mathbf{P}_e(\mathbf{r},t) = \sum_i \mathbf{P}_{e,i}^{co}(\mathbf{r} - \mathbf{u}_i(t),t)$ and $\mu_0^{-1}\mathbf{P}_m(\mathbf{r},t) = \sum_i \mathbf{P}_{e,i}^{co}(\mathbf{r} - \mathbf{u}_i(t),t) \times \mathbf{v}_i$ is the magnetization vector.

Generalizing Eq. 3.12 of Ref. [30] to the continuous limit and to the case of rigid bodies, we find that

$$\frac{d\mathbf{u}_i}{dt} = \mathbf{v}_i, \qquad (A2a)$$

$$\frac{d\mathbf{p}_{can,i}}{dt} = \mathbf{F}_{tot,i}^{ext} + \nabla_{\mathbf{u}_i} \int \mathbf{P}_{e,i}^{co}(\mathbf{R},t) \cdot \left[ \mathbf{E}(\mathbf{R}+\mathbf{u}_i,t) + \mathbf{v}_i \times \mathbf{B}(\mathbf{R}+\mathbf{u}_i,t) \right] d^3\mathbf{R}, \qquad (A2b)$$

where $\mathbf{p}_{can,i}$ is the canonical momentum of the $i$-th material body,

$$\begin{aligned}\mathbf{p}_{can,i} &= M_i\mathbf{v}_i - \int_{V_i} \mathbf{P}_e(\mathbf{r},t) \times \mathbf{B}(\mathbf{r},t)d^3\mathbf{r} \\ &= M_i\mathbf{v}_i - \int \mathbf{P}_{e,i}^{co}(\mathbf{R},t) \times \mathbf{B}(\mathbf{R}+\mathbf{u}_i,t)d^3\mathbf{R}\end{aligned} \quad (A3)$$

$M_i$ is the total mass of the body, $\mathbf{F}_{tot,i}^{ext}$ is an hypothetical external force acting on the moving body, and $V_i$ is the volumetric region occupied by the body in the lab frame (which may change with time). In the limit of no material dispersion (when the reduced mass of the moving dipoles is negligible) we can write

$$\mathbf{P}_{e,i}^{co}(\mathbf{R},t) = \varepsilon_0 \chi_{e,i}^{co}(\mathbf{R})\left[\mathbf{E}(\mathbf{R}+\mathbf{u}_i,t) + \mathbf{v}_i \times \mathbf{B}(\mathbf{R}+\mathbf{u}_i,t)\right], \quad (A4)$$

where $\chi_{e,i}^{co} > 0$ is the electric susceptibility of the moving body in its rest frame. Note that $\mathbf{E}+\mathbf{v}_i \times \mathbf{B}$ is the electric field in the co-moving frame.

Equations (A1)-(A4) completely determine the dynamics of the fields and of the moving bodies. Straightforward manipulations of Eqs. (A1) show that:

$$\begin{aligned}&\nabla \cdot (\mathbf{E} \times \mathbf{H}) + \partial_t W_{EM,P} + \mathbf{E} \cdot \mathbf{j}_{ext} = \\ &\frac{1}{2}\sum_i \left[\mathbf{P}_{e,i}^{co} \cdot (\partial_t \mathbf{E} + \mathbf{v}_i \times \partial_t \mathbf{B}) - \frac{d\mathbf{P}_{e,i}^{co}}{dt} \cdot (\mathbf{E}+\mathbf{v}_i \times \mathbf{B}) + \frac{d\mathbf{v}_i}{dt} \cdot (\mathbf{P}_{e,i}^{co} \times \mathbf{B})\right],\end{aligned} \quad (A5)$$

where $W_{EM,P} = \frac{1}{2}\mathbf{D}\cdot\mathbf{E} + \frac{1}{2}\mathbf{B}\cdot\mathbf{H}$ is the wave energy density. From Eq. (A4), $\mathbf{P}_{e,i}^{co}(\mathbf{r}-\mathbf{u}_i,t) = \varepsilon_0 \chi_{e,i}^{co}(\mathbf{r}-\mathbf{u}_i)\left[\mathbf{E}(\mathbf{r},t)+\mathbf{v}_i \times \mathbf{B}(\mathbf{r},t)\right]$ and hence we can also write:

$$\begin{aligned}&\nabla \cdot (\mathbf{E} \times \mathbf{H}) + \partial_t W_{EM,P} + \mathbf{E} \cdot \mathbf{j}_{ext} = \\ &\sum_i \left[\mathbf{v}_i \cdot \nabla_\mathbf{r} \chi_{e,i}^{co}(\mathbf{r}-\mathbf{u}_i) \frac{1}{2}\varepsilon_0 |\mathbf{E}+\mathbf{v}_i \times \mathbf{B}|^2 + \frac{d\mathbf{v}_i}{dt} \cdot (\mathbf{P}_{e,i}^{co} \times \mathbf{B})\right].\end{aligned} \quad (A6)$$

Thus, integrating over any volume $V$ that completely contains all the bodies (so that $\chi_{e,i}^{co}$ vanishes at the boundary $\partial V$ of the volume) we find after integration by parts:

$$\int_{\partial V} \hat{\mathbf{n}} \cdot (\mathbf{E} \times \mathbf{H}) ds + \frac{d}{dt} \int_V W_{EM,P} d^3\mathbf{r} + \int_V \mathbf{E} \cdot \mathbf{j}_{ext} d^3\mathbf{r}$$

$$= \sum_i \left[ -\int_V \mathbf{P}_{e,i}^{co} \cdot (\mathbf{v}_i \cdot \nabla_{\mathbf{r}})(\mathbf{E} + \mathbf{v}_i \times \mathbf{B}) d^3\mathbf{r} + \frac{d\mathbf{v}_i}{dt} \cdot \int_V \mathbf{P}_{e,i}^{co} \times \mathbf{B} d^3\mathbf{r} \right]$$

$$= \sum_i \left[ -\mathbf{v}_i \cdot \nabla_{\mathbf{u}_i} \int \mathbf{P}_e^{co}(\mathbf{R}) \cdot \left[ \mathbf{E}(\mathbf{R} + \mathbf{u}_i) + \mathbf{v}_i \times \mathbf{B}(\mathbf{R} + \mathbf{u}_i) \right] d^3\mathbf{r} + \frac{d\mathbf{v}_i}{dt} \cdot \int \mathbf{P}_{e,i}^{co}(\mathbf{R}) \times \mathbf{B}(\mathbf{R} + \mathbf{u}_i) d^3\mathbf{R} \right]$$

(A7)

where $\hat{\mathbf{n}}$ is the outward unit vector normal to the surface. But straightforward manipulations of Eqs. (A2) and (A3) show that the right hand side of the previous formula equals $-\frac{d}{dt} \left[ \sum_i \left( \mathbf{p}_{can,i} - \frac{1}{2} M_i \mathbf{v}_i \right) \cdot \mathbf{v}_i \right] + \sum_i \mathbf{F}_{tot,i}^{ext} \cdot \mathbf{v}_i$. Therefore, we demonstrated the energy conservation law:

$$\int_{\partial V} \hat{\mathbf{n}} \cdot (\mathbf{E} \times \mathbf{H}) ds + \frac{d}{dt} \int_V W_{EM,P} d^3\mathbf{r} + \frac{d}{dt} \left[ \sum_i \left( \mathbf{p}_{can,i} - \frac{1}{2} M_i \mathbf{v}_i \right) \cdot \mathbf{v}_i \right] = \sum_i \mathbf{F}_{tot,i}^{ext} \cdot \mathbf{v}_i - \int_V \mathbf{E} \cdot \mathbf{j}_{ext} d^3\mathbf{r}.$$

(A8)

In case $V$ is taken as all space, this result reduces to Eq. (3), being $H_{tot}$ defined as in Eq. (1) the total energy of the system. It interesting to note that the energy stored in the $i$-th body can be identified with:

$$H_{tot,i} = \mathbf{v}_i \cdot \left( \mathbf{p}_{can,i} - \frac{M_i \mathbf{v}_i}{2} \right) + \int_{V_i} W_{EM,P} d^3\mathbf{r}$$

$$= \frac{1}{2} M_i \mathbf{v}_i \cdot \mathbf{v}_i + \frac{1}{2} \int_{V_i} \varepsilon_0 \mathbf{E} \cdot \mathbf{E} + \mu_0^{-1} \mathbf{B} \cdot \mathbf{B} \, d^3\mathbf{r} + \mathbf{v}_i \cdot (\mathbf{p}_{can,i} - M_i \mathbf{v}_i) + \frac{1}{2} \int \mathbf{E} \cdot \mathbf{P}_{e,i}^{co} - \mathbf{B} \cdot (\mathbf{P}_{e,i}^{co} \times \mathbf{v}_i) d^3\mathbf{R}$$

(A9)

Using Eqs. (A3) and (A4) this can also be written as:

$$H_{tot,i} = \frac{1}{2}M_i \mathbf{v}_i \cdot \mathbf{v}_i + \frac{1}{2}\int_{V_i} \varepsilon_0 \mathbf{E} \cdot \mathbf{E} + \mu_0^{-1}\mathbf{B} \cdot \mathbf{B} \, d^3\mathbf{r} + \frac{1}{2}\int \frac{1}{\varepsilon_0 \chi_{e,i}^{co}} \mathbf{P}_{e,i}^{co} \cdot \mathbf{P}_{e,i}^{co} d^3\mathbf{R} \geq 0. \quad (A10)$$

This confirms that the energy stored in the $i$-th body is always positive, even in presence of system instabilities. Moreover, this result also implies that $H_{tot} \geq 0$.

Up to now the formalism is completely general. Next, we analyze the situation wherein the relevant bodies move along the $x$-direction ($\mathbf{v}_i = v_i \hat{\mathbf{x}}$), and are invariant to translations along $x$ ($\partial_x \chi_{e,i}^{co} = 0$). In such a case, substituting Eq. (A4) into (A2b), we get:

$$\begin{aligned}
\frac{dp_{can,i}}{dt} &= F_{i,x}^{ext} + \int \mathbf{P}_{e,i}^{co}(\mathbf{r}-\mathbf{u}_i,t) \cdot \frac{\partial}{\partial x}\left[\mathbf{E}(\mathbf{r},t) + \mathbf{v}_i \times \mathbf{B}(\mathbf{r},t)\right] d^3\mathbf{r} \\
&= F_{i,x}^{ext} + \int \varepsilon_0 \chi_{e,i}^{co}(\mathbf{r}-\mathbf{u}_i)\left[\mathbf{E}(\mathbf{r},t) + \mathbf{v}_i \times \mathbf{B}(\mathbf{r},t)\right] \cdot \frac{\partial}{\partial x}\left[\mathbf{E}(\mathbf{r},t) + \mathbf{v}_i \times \mathbf{B}(\mathbf{r},t)\right] d^3\mathbf{r} \\
&= F_{i,x}^{ext} + \int \frac{\partial}{\partial x}\left[\frac{1}{2}\mathbf{P}_{e,i}^{co}(\mathbf{r}-\mathbf{u}_i,t) \cdot \left(\mathbf{E}(\mathbf{r},t) + \mathbf{v}_i \times \mathbf{B}(\mathbf{r},t)\right)\right] d^3\mathbf{r} \\
&= F_{i,x}^{ext} + \int \frac{\partial}{\partial x}\left[\frac{1}{2}\mathbf{P}_e(\mathbf{r},t) \cdot \left(\mathbf{E}(\mathbf{r},t) + \mathbf{v}_i \times \mathbf{B}(\mathbf{r},t)\right)\right] d^3\mathbf{r} = F_{i,x}^{ext}
\end{aligned} \quad (A11)$$

where $p_{can,i}$ is the $x$-component of the canonical momentum and $F_{i,x}^{ext}$ is the $x$-component of the force. The last identity is obtained by assuming that the fields satisfy periodic boundary conditions in the spatial domain of interest. Thus, consistent with Ref. [17] we find that $\frac{dp_{can,i}}{dt} = F_{i,x}^{ext}$. Differentiating both sides of Eq. (A3) with respect to time, we can write:

$$\frac{dp_{kin,i}}{dt} = F_{i,x}^{ext} + \frac{d}{dt}\int_{V_i} \left(\mathbf{P}_e(\mathbf{r},t) \times \mathbf{B}(\mathbf{r},t)\right) \cdot \hat{\mathbf{x}} \, d^3\mathbf{r}. \quad (A12)$$

where $p_{kin,i} = M_i v_i$ is $x$-component of the kinetic momentum. The integral in the right-hand side can be identified with the pseudo-momentum introduced in the main text.

Indeed, in the non-relativistic limit and for bodies that only have an electric response in the rest frame, the pseudo-momentum satisfies:

$$\mathbf{p}_{ps,i} = \int_{V_i} \mathbf{P}_e \times \mathbf{B} + \varepsilon_0 \mathbf{E} \times \mathbf{P}_m d^3\mathbf{r}$$

$$= \int_{V_i} \mathbf{P}_e \times \mathbf{B} + \frac{1}{c} \mathbf{E} \times \left( \mathbf{P}_{e,i}^{co} \times \frac{\mathbf{v}_i}{c} \right) d^3\mathbf{r} . \quad (A13)$$

$$\approx \int_{V_i} \mathbf{P}_e \times \mathbf{B} d^3\mathbf{r}$$

The last identity follows from the fact that $v_i/c \ll 1$ and that $\mathbf{E}/c$ can be estimated to be of the same order of magnitude as $\mathbf{B}$. Thus, Eq. (A12) and Eq. (A13) imply that the *x*-component of the momentum can be decomposed as in Eq. (4) of the main text.

## Appendix B:

Here, we compute the contribution from a generic term of the series (14) to the friction force. For convenience, we introduce the following bilinear forms:

$$p_{EM,i}(\mathbf{F}_1, \mathbf{F}_2) = \frac{1}{2} \frac{1}{c^2} \int_{V_i} d^3\mathbf{r} \left( \mathbf{E}_1 \times \mathbf{H}_2 + \mathbf{E}_2 \times \mathbf{H}_1 \right) \cdot \hat{\mathbf{x}}, \quad (B1a)$$

$$p_{wv,i}(\mathbf{F}_1, \mathbf{F}_2) = \frac{1}{2} \int_{V_i} d^3\mathbf{r} \left( \mathbf{D}_1 \times \mathbf{B}_2 + \mathbf{D}_2 \times \mathbf{B}_1 \right) \cdot \hat{\mathbf{x}}, \quad (B1b)$$

$$p_{ps,i}(\mathbf{F}_1, \mathbf{F}_2) = p_{wm,i}(\mathbf{F}_1, \mathbf{F}_2) - p_{EM,i}(\mathbf{F}_1, \mathbf{F}_2). \quad (B1c)$$

Evidently, the electromagnetic momentum, the wave momentum, and the pseudo-momentum stored in the *i*-th slab for a real-valued field distribution $\mathbf{F}$ are $p_{EM,i} = p_{EM,i}(\mathbf{F},\mathbf{F})$, $p_{wm,i} = p_{wm,i}(\mathbf{F},\mathbf{F})$ and $p_{ps,i} = p_{ps,i}(\mathbf{F},\mathbf{F})$, respectively. Let us first consider a generic term of the series (14a) associated with real-valued frequencies of oscillation. Because the spatial variation along *x* and *y* is of the form $e^{i\mathbf{k}\cdot\mathbf{r}}$ it is simple to check that $p_{u,i}(\mathbf{F}_{n\mathbf{k}}, \mathbf{F}_{n\mathbf{k}}) = 0 = p_{u,i}(\mathbf{F}_{n\mathbf{k}}^*, \mathbf{F}_{n\mathbf{k}}^*)$, with *u=EM, wv, ps*. Hence, it follows that:

$$p_{u,i} = \frac{\hbar |\omega_{n\mathbf{k}}|}{2} \left[ \hat{c}_{n\mathbf{k}} \hat{c}_{n\mathbf{k}}^\dagger p_{u,i}\left(\mathbf{F}_{n\mathbf{k}}, \mathbf{F}_{n\mathbf{k}}^*\right) + \hat{c}_{n\mathbf{k}}^\dagger \hat{c}_{n\mathbf{k}} p_{u,i}\left(\mathbf{F}_{n\mathbf{k}}^*, \mathbf{F}_{n\mathbf{k}}\right) \right]. \tag{B2}$$

Thus, $p_{u,i}$ is independent of time, and so it does not contribute to the friction force $dp_{u,i}/dt = 0$.

Next, we consider a generic term of the series (14b), $\hat{\mathbf{F}}_C = \hat{\beta}_{n\mathbf{k}} e^{-i\omega_{c,n\mathbf{k}} t} \mathbf{f}_{n\mathbf{k}} + \hat{\chi}_{n\mathbf{k}} e^{-i\omega_{c,n\mathbf{k}}^* t} \mathbf{e}_{n\mathbf{k}} + \hat{\beta}_{n\mathbf{k}}^\dagger e^{i\omega_{c,n\mathbf{k}}^* t} \mathbf{f}_{n\mathbf{k}}^* + \hat{\chi}_{n\mathbf{k}}^\dagger e^{i\omega_{c,n\mathbf{k}} t} \mathbf{e}_{n\mathbf{k}}^*$, associated with a pair of complex-valued frequencies of oscillation. Proceeding as in the previous case, it is straightforward to prove that:

$$\begin{aligned}
p_{u,i} &= p_{u,i}\left( \hat{\beta}_{n\mathbf{k}} e^{-i\omega_{c,n\mathbf{k}} t} \mathbf{f}_{n\mathbf{k}} + \hat{\chi}_{n\mathbf{k}} e^{-i\omega_{c,n\mathbf{k}}^* t} \mathbf{e}_{n\mathbf{k}}, \hat{\beta}_{n\mathbf{k}}^\dagger e^{i\omega_{c,n\mathbf{k}}^* t} \mathbf{f}_{n\mathbf{k}}^* + \hat{\chi}_{n\mathbf{k}}^\dagger e^{i\omega_{c,n\mathbf{k}} t} \mathbf{e}_{n\mathbf{k}}^* \right) \\
&+ p_{u,i}\left( \hat{\beta}_{n\mathbf{k}}^\dagger e^{i\omega_{c,n\mathbf{k}}^* t} \mathbf{f}_{n\mathbf{k}}^* + \hat{\chi}_{n\mathbf{k}}^\dagger e^{i\omega_{c,n\mathbf{k}} t} \mathbf{e}_{n\mathbf{k}}^*, \hat{\beta}_{n\mathbf{k}} e^{-i\omega_{c,n\mathbf{k}} t} \mathbf{f}_{n\mathbf{k}} + \hat{\chi}_{n\mathbf{k}} e^{-i\omega_{c,n\mathbf{k}}^* t} \mathbf{e}_{n\mathbf{k}} \right)
\end{aligned} \tag{B3}$$

Thus, using $\omega_{c,n\mathbf{k}} = \omega_{n\mathbf{k}}' + i\lambda_{n\mathbf{k}}$ one sees that the time derivative of the pertinent momentum ($u$=EM, wv, ps) is given by:

$$\begin{aligned}
\frac{dp_{u,i}}{dt} &= \frac{d}{dt}\left[ e^{2\lambda_{n\mathbf{k}} t} \hat{\beta}_{n\mathbf{k}} \hat{\beta}_{n\mathbf{k}}^\dagger p_{u,i}\left(\mathbf{f}_{n\mathbf{k}}, \mathbf{f}_{n\mathbf{k}}^*\right) + e^{-2\lambda_{n\mathbf{k}} t} \hat{\chi}_{n\mathbf{k}} \hat{\chi}_{n\mathbf{k}}^\dagger p_{u,i}\left(\mathbf{e}_{n\mathbf{k}}, \mathbf{e}_{n\mathbf{k}}^*\right) \right] \\
&+ \frac{d}{dt}\left[ e^{2\lambda_{n\mathbf{k}} t} \hat{\beta}_{n\mathbf{k}}^\dagger \hat{\beta}_{n\mathbf{k}} p_{u,i}\left(\mathbf{f}_{n\mathbf{k}}^*, \mathbf{f}_{n\mathbf{k}}\right) + e^{-2\lambda_{n\mathbf{k}} t} \hat{\chi}_{n\mathbf{k}}^\dagger \hat{\chi}_{n\mathbf{k}} p_{u,i}\left(\mathbf{e}_{n\mathbf{k}}^*, \mathbf{e}_{n\mathbf{k}}\right) \right]
\end{aligned} \tag{B4}$$

From the definition of the bilinear forms (B1) $p_{u,i}\left(\mathbf{f}_{n\mathbf{k}}, \mathbf{f}_{n\mathbf{k}}^*\right) = p_{u,i}\left(\mathbf{f}_{n\mathbf{k}}^*, \mathbf{f}_{n\mathbf{k}}\right)$ is real-valued.

Moreover, using Eq. (13) in Eq. (B1a) it is found that if $\mathbf{f} = (\mathbf{E} \quad \mathbf{H})^T$ then:

$$p_{EM,i}\left(\mathbf{e}, \mathbf{e}^*\right) = \frac{1}{2}\int_{V_i} d^3\mathbf{r} \left[ \left(\mathbf{R}_{z,\pi}\mathbf{E}^*\right)\times\left(-\mathbf{R}_{z,\pi}\mathbf{H}\right) + \left(\mathbf{R}_{z,\pi}\mathbf{E}\right)\times\left(-\mathbf{R}_{z,\pi}\mathbf{H}^*\right) \right] \cdot \hat{\mathbf{x}}. \tag{B5}$$

But for generic vectors $\mathbf{v}_1$ and $\mathbf{v}_2$, one has $\left(\mathbf{R}_{z,\pi}\mathbf{v}_1\right)\times\left(-\mathbf{R}_{z,\pi}\mathbf{v}_2\right) = -\mathbf{R}_{z,\pi}\cdot\left(\mathbf{v}_1\times\mathbf{v}_2\right)$, and hence:

$$p_{EM,i}(\mathbf{e},\mathbf{e}^*) = \frac{1}{2c^2}\int_{V_i} d^3\mathbf{r}\left(\mathbf{E}^* \times \mathbf{H} + \mathbf{E} \times \mathbf{H}^*\right)\cdot\hat{\mathbf{x}} = p_{EM,i}(\mathbf{f},\mathbf{f}^*). \tag{B6}$$

On the other hand, Eq. (13) and the properties $\mathbf{M}(\mathbf{r}) = \mathbf{M}(\mathbf{R}_{z,\pi}\cdot\mathbf{r})$ and $\mathbf{U}\cdot\mathbf{U} = \mathbf{1}$, imply that $\tilde{\mathbf{g}}(\mathbf{r}) \equiv \mathbf{M}(\mathbf{r})\cdot\tilde{\mathbf{f}}(\mathbf{r})$ can be written as $\tilde{\mathbf{g}}(\mathbf{r}) = \mathbf{U}\cdot\mathbf{U}\cdot\mathbf{M}(\mathbf{R}_{z,\pi}\cdot\mathbf{r})\cdot\mathbf{U}\cdot\mathbf{f}^*(\mathbf{R}_{z,\pi}\cdot\mathbf{r})$. Taking into account the explicit form of the material matrix $\mathbf{M}$, which can be found in Ref. [17], it is possible to show that because our system is invariant to a 180º rotation around the z-axis, $\mathbf{M} = \mathbf{U}\cdot\mathbf{M}\cdot\mathbf{U}$. Thus, this shows that $\tilde{\mathbf{g}}(\mathbf{r}) = \mathbf{U}\cdot\mathbf{g}^*(\mathbf{R}_{z,\pi}\cdot\mathbf{r})$. Therefore, the $\mathbf{D}$ and $\mathbf{B}$ fields associated with $\mathbf{e} = \tilde{\mathbf{f}}$ are related to the $\mathbf{D}$ and $\mathbf{B}$ fields associated with $\mathbf{f}$, in the same manner as the $\mathbf{E}$ and $\mathbf{H}$ fields associated with $\mathbf{e} = \tilde{\mathbf{f}}$ are related to the $\mathbf{E}$ and $\mathbf{H}$ fields associated with $\mathbf{f}$. This implies that:

$$p_{wm,i}(\mathbf{e},\mathbf{e}^*) = \frac{1}{2}\int_{V_i} d^3\mathbf{r}\left(\mathbf{D}^* \times \mathbf{B} + \mathbf{D} \times \mathbf{B}^*\right)\cdot\hat{\mathbf{x}} = p_{wm,i}(\mathbf{f},\mathbf{f}^*). \tag{B7}$$

Equations (B6)-(B7) also make obvious that $p_{ps,i}(\mathbf{e},\mathbf{e}^*) = p_{ps,i}(\mathbf{f},\mathbf{f}^*)$. Substituting these results into Eq. (B4) we find that:

$$\frac{dp_{u,i}}{dt} = 2\lambda_{n\mathbf{k}} p_{u,i}(\mathbf{f}^*_{n\mathbf{k}},\mathbf{f}_{n\mathbf{k}})\left[e^{2\lambda_{n\mathbf{k}}t}\left(\hat{\beta}_{n\mathbf{k}}\hat{\beta}^\dagger_{n\mathbf{k}} + \hat{\beta}^\dagger_{n\mathbf{k}}\hat{\beta}_{n\mathbf{k}}\right) - e^{-2\lambda_{n\mathbf{k}}t}\left(\hat{\chi}_{n\mathbf{k}}\hat{\chi}^\dagger_{n\mathbf{k}} + \hat{\chi}^\dagger_{n\mathbf{k}}\hat{\chi}_{n\mathbf{k}}\right)\right]. \tag{B8}$$

In the framework of the quantum theory, $\hat{\beta}_{n\mathbf{k}}$ and $\hat{\chi}_{n\mathbf{k}}$ are operators. Using $\left[\hat{\beta}_{n\mathbf{k}},\hat{\beta}^\dagger_{n\mathbf{k}}\right] = 0 = \left[\hat{\chi}_{n\mathbf{k}},\hat{\chi}^\dagger_{n\mathbf{k}}\right]$ and the decomposition (16) one can write:

$$\begin{aligned}\frac{dp_{u,i}}{dt} = &\hbar\left|\omega_{c,n\mathbf{k}}\right|\lambda_{n\mathbf{k}} p_{u,i}(\mathbf{f}^*_{n\mathbf{k}},\mathbf{f}_{n\mathbf{k}})\\&\times\left[e^{2\lambda_{n\mathbf{k}}t}\left(\hat{a}_{c,n\mathbf{k}} + \hat{b}^\dagger_{c,n\mathbf{k}}\right)\left(\hat{a}^\dagger_{c,n\mathbf{k}} + \hat{b}_{c,n\mathbf{k}}\right) - e^{-2\lambda_{n\mathbf{k}}t}\left(\hat{a}_{c,n\mathbf{k}} - \hat{b}^\dagger_{c,n\mathbf{k}}\right)\left(\hat{a}^\dagger_{c,n\mathbf{k}} - \hat{b}_{c,n\mathbf{k}}\right)\right]\end{aligned}. \tag{B9}$$

After straightforward simplifications, we finally obtain:

$$\frac{dp_{u,i}}{dt} = \hbar |\omega_{c,n\mathbf{k}}| 2\lambda_{n\mathbf{k}} p_{u,i} \left(\mathbf{f}_{n\mathbf{k}}^*, \mathbf{f}_{n\mathbf{k}}\right)$$
$$\times \left[ \left(\hat{a}_{c,n\mathbf{k}} \hat{a}_{c,n\mathbf{k}}^\dagger + \hat{b}_{c,n\mathbf{k}}^\dagger \hat{b}_{c,n\mathbf{k}}\right) \sinh 2\lambda_{n\mathbf{k}} t + \left(\hat{a}_{c,n\mathbf{k}} \hat{b}_{c,n\mathbf{k}} + \hat{b}_{c,n\mathbf{k}}^\dagger \hat{a}_{c,n\mathbf{k}}^\dagger\right) \cosh 2\lambda_{n\mathbf{k}} t \right]$$
(B10)

## Appendix C:

Here, we prove that in the limit of a weak slab interaction and for $v_i/c \ll 1$ it is possible to write $\left|\hat{F}_i^{tot}\right|$ in terms of the imaginary part of the reflection coefficients, establishing in this manner a connection with the original theory of Pendry [1, 4].

For $v_i/c \ll 1$ the characteristic equation for the natural modes of oscillation of the system depicted in Fig. 1a can be written as $D^p(\omega, k_x, k_y) D^s(\omega, k_x, k_y) = 0$ being $D^l(\omega, k_x, k_y)$ given by [18]

$$D^l(\omega, k_x, k_y) = 1 - e^{-2\gamma_0 d} R_{co,1}^l(\omega - v_1 k_x, k_x, k_y) R_{co,2}^l(\omega - v_2 k_x, k_x, k_y), \quad l=s,p \quad (C1)$$

where $p$ and $s$ specify the wave polarization, and $d$ is the distance between the two interacting bodies. In the above, $R_{co}^p(\tilde{\omega}, k_x, k_y)$ and $R_{co}^s(\tilde{\omega}, k_x, k_y)$ represent the reflection coefficients calculated in the pertinent co-moving frame, supposing that a plane wave propagating in vacuum with transverse wave vector $(k_x, k_y)$ illuminates the pertinent slab. For the particular geometry of Fig. 1a, one has:

$$R_{co}^l(\omega, k_x, k_y) = \frac{Y_0^l - \coth(\gamma_d h) Y_d^l}{Y_0^l + \coth(\gamma_d h) Y_d^l}, \quad l=s,p. \quad (C2)$$

where $Y^p = \dfrac{\varepsilon}{i\gamma} \dfrac{\omega}{c}$ and $Y^s = \dfrac{c}{\omega \mu} i\gamma$ are the normalized wave admittances for $p$ and $s$ polarized waves. In the above, $\varepsilon$ and $\mu$ are the relative permittivity and permeability, respectively, and $\gamma = \sqrt{k_x^2 + k_y^2 - \varepsilon\mu\omega^2/c^2}$ is the propagation constant along the $+z$ direction. The subscripts "*0*" and "*d*" indicate if $\varepsilon$ and $\mu$ are either calculated in the vacuum region ($\varepsilon = \mu = 1$) or in dielectric region ($\varepsilon = \varepsilon_d$ and $\mu = \mu_d$).

To make further progress, we note that for a weak interaction the coupling term $e^{-2\gamma_0 d} \approx e^{-2kd}$ in Eq. (C1) is vanishingly small. Here, we neglect the effects of time retardation in the vacuum layer, and thus $\gamma_0 \approx k \equiv \sqrt{k_x^2 + k_y^2}$. Hence, possible solutions of $D^l(\omega, k_x, k_y) = 0$ necessarily occur in the vicinity of the poles of $R_{co,i}^l$. The poles of $R_{co,i}^l$ determine the guided modes of the $i$-th slab.

We are interested in the modes of oscillation with $\omega$ complex-valued. The poles of $R_{co,i}^l$ are real-valued in the absence of material loss in the dielectrics. Moreover, because the complex zeros of $D^l$ necessarily occur in pairs $\omega = \omega' \pm i\lambda$, they must be the result of the interaction of two poles of $R_{co,1}^l R_{co,2}^l$. In other words, the emergence of complex zeros $\omega = \omega' \pm i\lambda$ requires that $R_{co,1}^l R_{co,2}^l$ has two closely spaced poles, because otherwise $D^l = 0$ has a single real valued zero. Supposing for simplicity that $R_{co,i}^l$ has only simple poles (of order one) it follows that system instabilities are a consequence of the interaction of a pole of $R_{co,1}^l$ and a pole of $R_{co,2}^l$, or in other words of the interaction of a guided mode of the first slab and a guided mode of the second slab.

For a fixed $(k_x, k_y)$, let $\tilde{\omega}_1$ be a pole of $R_{co,1}^l(\tilde{\omega}, k_x, k_y)$ and $\tilde{\omega}_2$ be a pole of $R_{co,2}^l(\tilde{\omega}, k_x, k_y)$. In the vicinity of the poles, we can write:

$$R_{co,i}^l(\tilde{\omega}, k_x, k_y) \approx \frac{b_i}{\tilde{\omega} - \tilde{\omega}_i} \tag{C3}$$

where $b_i$ is the residue of the pole. Taking into account the Doppler shifts, one sees that in order that these two poles can generate a natural mode with complex-valued $\omega = \omega' \pm i\lambda$, it is necessary that $\omega_1 \approx \omega_2$ where $\omega_1 = \tilde{\omega}_1 + v_1 k_x$ and $\omega_2 = \tilde{\omega}_2 + v_2 k_x$ are the Doppler shifted poles. It is interesting to note that the condition that the two Doppler-shifted poles are closely spaced, $\omega_1 \approx \omega_2$, yields a "selection rule" for $k_x$:

$$k_x = -\frac{\tilde{\omega}_2 - \tilde{\omega}_1}{v_2 - v_1}. \tag{C4}$$

One should keep in mind that in general both $\tilde{\omega}_1$ and $\tilde{\omega}_2$ may depend on $(k_x, k_y)$. Only poles $\tilde{\omega}_1$ and $\tilde{\omega}_2$ that satisfy approximately this selection rule can generate modes of oscillation with $\lambda \neq 0$. Moreover, to have system instabilities a second selection rule needs to be satisfied

$$\tilde{\omega}_1 \tilde{\omega}_2 < 0, \tag{C5}$$

i.e. the poles must be associated with frequencies (as calculated in the respective co-moving frames) with different signs. To derive this second selection rule, we note that for $\omega$ in the vicinity of $\omega_1 \approx \omega_2$ the characteristic equation is simplified to:

$$D^l(\omega, k_x, k_y) = 1 - e^{-2kd} \frac{b_1}{\omega - \omega_1} \frac{b_2}{\omega - \omega_2} \tag{C6}$$

To calculate $\lambda$, we solve this equation with respect to $\omega$:

$$\omega = \frac{\omega_1 + \omega_2}{2} \pm \sqrt{e^{-2kd} b_1 b_2 + \left(\frac{\omega_1 - \omega_2}{2}\right)^2}. \tag{C7}$$

Thus, in order to have complex-valued solutions $\omega = \omega' + i\lambda$ it is necessary that the residues satisfy:

$$b_1 b_2 < 0. \tag{C8a}$$

$$|\omega_1 - \omega_2| < 2\lambda_0, \qquad \text{with} \quad \lambda_0 = e^{-kd}|b_1 b_2|^{1/2}. \tag{C8b}$$

It can be checked using Eq. (C2) that in the lossless limit the sign of the residue $b_i$ is strictly linked to the sign of the corresponding pole $\tilde{\omega}_i$, such that the condition $b_1 b_2 < 0$ is equivalent to $\tilde{\omega}_1 \tilde{\omega}_2 < 0$. This justifies the second selection rule [Eq. (C5)]. On the other hand, in the limit of a weak interaction condition (C8b) implies that $\omega_1 \approx \omega_2$, and thus it provides a more strict formulation of the first selection rule [Eq. (C4)].

For complex-valued poles the imaginary part $\lambda$ satisfies:

$$\begin{aligned}\lambda &= \sqrt{e^{-2kd}|b_1 b_2| - \left(\frac{\omega_1 - \omega_2}{2}\right)^2} \\ &= \lambda_0 \sqrt{1 - \left(\frac{\omega_1 - \omega_2}{2\lambda_0}\right)^2}\end{aligned}. \tag{C9}$$

Because $\omega_i = \tilde{\omega}_i(k_x, k_y) + v_i k_x$ and $b_i = b_i(k_x, k_y)$, $i=1,2$, depend on $k_x, k_y$ we can regard $\lambda$ as a function of $k_x, k_y$. From the previous discussion, it can be seen that $\lambda(k_x, k_y) = 0$, except in the region of the $(k_x, k_y)$ plane wherein the conditions (C8) are simultaneously satisfied. For a weak interaction, this (two-dimensional) region is certainly confined to the close proximity of the (one-dimensional) curve wherein the selection rule (C4) is exactly satisfied, i.e. wherein $\omega_1 = \omega_2$. Thus, we can approximate $\lambda(k_x, k_y)$ by:

$$\lambda(k_x, k_y) \approx B\delta(\omega_2 - \omega_1). \tag{C10}$$

where $B = B(k_x, k_y)$ is some function to be determined. Specifically, we impose that $B$ is such that $\iint \lambda(k_x, k_y) dk_x dk_y$ is the same, independent of one using the exact formula for $\lambda$ [Eq. (C9)] or the approximate formula [Eq. (C10)], similar to what has been done in Ref. [8] for a related problem.

Let $\left(k_x^0, k_y^0\right)$ be some point lying in the curve determined by (C4), and let us consider a small interval of area $\Delta k_x \times \Delta k_y$ centered at this point. The integral of $\lambda\left(k_x, k_y\right)$ calculated over this interval using (C10) gives:

$$\iint \lambda\left(k_x, k_y\right) dk_x dk_y \approx \frac{B}{\left|\partial_{k_x}\omega_2 - \partial_{k_x}\omega_1\right|} \Delta k_y . \tag{C11}$$

where the right-hand side is evaluated at $\left(k_x^0, k_y^0\right)$. On the other hand, the same integral calculated using Eq. (C9) gives:

$$\begin{aligned}\iint \lambda\left(k_x, k_y\right) dk_x dk_y &\approx \lambda_0 \iint \sqrt{1 - \left(\frac{\omega_1 - \omega_2}{2\lambda_0}\right)^2} dk_x dk_y \\ &\approx \lambda_0 \iint \sqrt{1 - \left(\frac{\partial_{k_x}\omega_1 - \partial_{k_x}\omega_2}{2\lambda_0}\right)^2 \left(k_x - k_x^0\right)^2} dk_x dk_y \\ &= \frac{\pi}{2} \frac{1}{\left|\frac{\partial_{k_x}\omega_1 - \partial_{k_x}\omega_2}{2\lambda_0}\right|} \lambda_0 \int dk_y = \frac{\pi \lambda_0^2}{\left|\partial_{k_x}\omega_1 - \partial_{k_x}\omega_2\right|} \Delta k_y\end{aligned} \tag{C12}$$

where we used $\int \sqrt{1 - A^2\left(k_x - k_x^0\right)^2} dk_x = \frac{\pi}{2} \frac{1}{|A|}$, and the rightmost expression is evaluated at $\left(k_x^0, k_y^0\right)$. Comparing the results of the integration in Eqs. (C11)-(C12) it follows that $B = \pi \lambda_0^2$, and substituting this result into Eq. (C10) we find that:

$$\begin{aligned}\lambda_{\mathbf{k}} &\approx \pi e^{-2kd} \left|b_1 b_2\right| \delta\left(\omega_2 - \omega_1\right) \\ &= -\pi e^{-2kd} \int d\omega\, b_1 \delta\left(\omega - \omega_1\right) b_2 \delta\left(\omega - \omega_2\right)\end{aligned} . \tag{C13}$$

In the second identity we used the fact that $b_1 b_2 < 0$, and it is implicit that only a pair of poles with $\tilde{\omega}_1 \tilde{\omega}_2 < 0$ gives a nonzero $\lambda$. From Eq. (C3) one has $\text{Im}\{R^l_{co,i}(\tilde{\omega}, k_x, k_y)\} \approx -\pi b_i \delta(\tilde{\omega} - \tilde{\omega}_i)$ in the vicinity of $\tilde{\omega} = \tilde{\omega}_i$. Thus, one can take into account the interaction of *all poles* of $R^l_{co,1}(\tilde{\omega}, k_x, k_y)$ and $R^l_{co,2}(\tilde{\omega}, k_x, k_y)$ simply by replacing Eq. (C13) [which takes into account the contribution of a single pair of poles] by:

$$\lambda_{\mathbf{k}} = -\frac{1}{\pi} e^{-2kd} \int_{(\omega - k_x v_1)(\omega - k_x v_2) < 0} d\omega \, \text{Im}\{R^l_{co,1}(\omega - k_x v_1, k_x, k_y)\} \text{Im}\{R^l_{co,2}(\omega - k_x v_2, k_x, k_y)\} \quad (C14)$$

The integration region was restricted to $(\omega - k_x v_1)(\omega - k_x v_2) < 0$ to ensure that only pairs of poles with $\tilde{\omega}_1 \tilde{\omega}_2 < 0$ are considered. This can also be rewritten as:

$$\lambda_{\mathbf{k}} = -\frac{1}{\pi} e^{-2kd} \, \text{sgn}(k_x(v_1 - v_2)) \int_{k_x v_2}^{k_x v_1} d\omega \, \text{Im}\{R^l_{co,1}(\omega - k_x v_1, k_x, k_y)\} \text{Im}\{R^l_{co,2}(\omega - k_x v_2, k_x, k_y)\}$$

(C15)

As expected, when $v_1 = v_2$ the above formula gives $\lambda_{\mathbf{k}} = 0$.

We are now ready to compute the friction force variance per unit of area when the system is in the pseudo-ground state. From Eqs. (26) and (C15), we can write:

$$\frac{|\hat{F}_i^{tot,l}|}{A_0} = -\text{sgn}(v_1 - v_2) \frac{\hbar}{4\pi^3} \int_{-\infty}^{+\infty} dk_y \int_0^{+\infty} dk_x \left[ k_x e^{-2kd} \right. \\ \left. \times \int_{k_x v_2}^{k_x v_1} d\omega \, \text{Im}\{R^l_{co,1}(\omega - k_x v_1, k_x, k_y)\} \text{Im}\{R^l_{co,2}(\omega - k_x v_2, k_x, k_y)\} \right]. \quad (C16)$$

This result is coincident with Pendry's friction formula when the contributions from multi-scattering are neglected [1, 4], as we wanted to prove (see in particular Eq. 12 of Ref. [4]; note that a factor of ½ was inserted by Pendry in Eq. 12 of Ref. [4] to correct the original theory of Ref. [1]).

It is interesting to note that if we substitute Eq. (C10) with $B = \pi \lambda_0^2 = \pi e^{-2kd} |b_1 b_2|$ into Eq. (26) we obtain:

$$\frac{\left|\hat{F}_i^{tot}\right|}{A_0} = \frac{\hbar}{(2\pi)^2} \int_{-\infty}^{+\infty} dk_y \, k_x \pi e^{-2kd} \frac{|b_1 b_2|}{\left|\partial_{k_x}\omega_1 - \partial_{k_x}\omega_2\right|} . \tag{C17}$$

where $(k_x, k_y)$ is a generic wave vector that satisfies the selection rules (C4) and (C5) with $k_x > 0$, and $\omega_i = \tilde{\omega}_i(k_x, k_y) + v_i k_x$ and $b_i = b_i(k_x, k_y)$ are defined as outlined previously. In the case of identical slabs, the dominant contribution to the friction force is due to branches with $\tilde{\omega}_2 = -\tilde{\omega}_1$. Neglecting all the other friction force channels, and assuming without loss of generality that $v_2 - v_1 > 0$, so that the condition $k_x > 0$ implies $\tilde{\omega}_1 > 0$, we see that the selection rules reduce to:

$$k_x = \frac{2\tilde{\omega}_1}{v_2 - v_1} \quad \text{and} \quad \tilde{\omega}_1 > 0. \tag{C18}$$

Using these results and $-b_1 b_2 = b_1^2$ in Eq. (C17), it is found that:

$$\frac{\left|\hat{F}_i^{tot}\right|}{A_0} = \frac{\hbar c}{(2\pi)^2} \int_{-\infty}^{+\infty} dk_y \, k_x \, \pi e^{-2kd} \frac{b_1^2 / c^2}{\left|2\frac{1}{c}\partial_{k_x}\tilde{\omega}_1 - \frac{1}{c}(v_2 - v_1)\right|} . \tag{C19}$$

This formula gives the friction force directly in terms of the dispersion of the guided modes of a single slab as seen in the co-moving frame ($\tilde{\omega}_1(k_x, k_y)$) and in terms of the associated residue of the reflection coefficient ($b_1 = b_1(k_x, k_y)$). The integration should be done over all the "curves" in the $(k_x, k_y)$ plane that satisfy Eq. (C18), for both *s*- and *p*-polarized waves.

## Appendix D:

Here, we estimate the relation between the forces $F_i^{tot}$ and $F_i^{mat}$ near the friction force threshold for the case of two moving dielectric bodies with refractive indices (in the respective co-moving frames) $n_1$ and $n_2$. The bodies have velocity $v_1$ and $v_2$ with respect to the lab frame, and are separated by a vacuum gap with thickness *d*, as in Fig. 1a.

To begin with, we calculate the wave- and pseudo-momentum of the moving bodies in the limit of a weak coupling. Let the mode **f** be associated with a complex-valued frequency of oscillation $\omega = \omega' + i\lambda$ with $\lambda > 0$ and the mode $\mathbf{e} = \tilde{\mathbf{f}}$ [Eq. (13)] be associated with $\omega = \omega' - i\lambda$. The natural modes are associated with the real-valued transverse wave vector $\mathbf{k} = (k_x, k_y, 0)$ [variation along *x* and *y* coordinates is of the form $e^{i\mathbf{k} \cdot \mathbf{r}}$]. As usual the modes **f** and **e** satisfy the normalization conditions $\langle \mathbf{e} | \mathbf{e} \rangle = \langle \mathbf{f} | \mathbf{f} \rangle = 0$ and $\langle \mathbf{e} | \mathbf{f} \rangle = 1$. Following Ref. [17, Eq. 24], the wave momentum of the *i*-th body (at *t=0*) satisfies:

$$p_{wv,i} \equiv p_{wv,i}(\mathbf{f}^*, \mathbf{f}) \approx \frac{1}{2} \frac{k_x}{\omega'} \text{sgn}(E_{s,i}) \approx \pm \frac{1}{2} \frac{k_x}{|\omega|}. \tag{D1}$$

where $\text{sgn}(E_{s,i}) = \pm 1$ depending on the sign of the wave energy $E_{s,i}$ stored in the *i*-th slab, and $p_{u,i}(.,.)$ is the bilinear form introduced in Appendix B [Eq. B1]. Note that in presence of wave instabilities $E_{s,1} E_{s,2} < 0$, and thus $p_{wv,1} = -p_{wv,2}$, i.e. the two bodies have symmetric wave momenta [17].

On the other hand, again from the results of Ref. [17, Eqs. (24)-(25)] it is possible to write:

$$p_{ps,i} \equiv p_{ps,i}\left(\mathbf{f}^*,\mathbf{f}\right) = p_{wv,i}\left(\mathbf{f}^*,\mathbf{f}\right)\left(1 - \frac{v_{ph,i} v_{g,i}}{c^2}\right)$$
$$= p_{wv,i}\left(\mathbf{f}^*,\mathbf{f}\right)\left(1 - \frac{1}{c^2} \frac{v_{ph,i}^{co} + v_i}{1 + v_{ph,i}^{co} v_i / c^2} \frac{v_{g,i}^{co} + v_i}{1 + v_{g,i}^{co} v_i / c^2}\right). \tag{D2}$$

In the above, $v_{ph,i} = \omega_i' / k_x$ and $v_{g,i} = \partial \omega_i' / \partial k_x$ are the phase and group velocities in the lab frame, $v_{ph,i}^{co}$ and $v_{g,i}^{co}$ are the corresponding parameters calculated in the co-moving frame, and $\omega_i' = \omega_i'(k_x, k_y, v_i)$ is understood as the dispersion of the guided mode supported by the $i$-th slab in the absence of interaction.

As discussed in Sect. IV, at the friction force threshold it is expected that the modes of the individual slabs that originate system instabilities have $k_y = 0$ and $n_{ph,i} \approx \pm n_i$, where $n_{ph,i} = ck / \tilde{\omega}_i$ is the "phase" refractive index for the $i$-th slab (calculated in the co-moving frame), being $k = \sqrt{k_x^2 + k_y^2}$. It is known that when $|n_{ph,i}| \approx n_i$ the group and phase velocities of the guided modes (for slabs made of nondispersive materials) are nearly identical: $v_{g,i}^{co} \approx v_{ph,i}^{co}$ (see Fig. 2a). Therefore, from Eq. (D2), it follows that (supposing that $n_i \gg 1$ so that $v_i / c \ll 1$):

$$p_{ps,i}\left(\mathbf{f}^*,\mathbf{f}\right) \approx p_{wv,i}\left(\mathbf{f}^*,\mathbf{f}\right)\left(1 - \frac{1}{c^2 k_x^2}(\tilde{\omega}_i + v_i k_x)^2\right). \tag{D3}$$

In particular, because of the selection rules (18) and because $p_{wv,1} = -p_{wv,2}$ it follows that:

$$p_{ps,1} = -p_{ps,2}, \tag{D4}$$

i.e. the two interacting slabs also have symmetric pseudo-momenta. Thus, provided the velocities of the slabs are not too far from the friction threshold (so that $|\tilde{\omega}_i / k_x| \approx c / n_i$.), we can write:

$$p_{ps,i} \approx p_{wv,i}\left(1 - \left(\frac{\pm 1}{n_i} + \frac{v_i}{c}\right)^2\right). \tag{D5}$$

Taking into account that $F_i^{tot}$ is given by Eq. (17) and $F_i^{mat}$ is given by a similar formula with $p_{wv,i}(\mathbf{f}_{n\mathbf{k}})$ replaced by $p_{ps,i}(\mathbf{f}_{n\mathbf{k}})$, we can also write $F_i^{mat} \approx F_i^{tot}\left(1 - \left(\frac{\pm 1}{n_i} + \frac{v_i}{c}\right)^2\right)$. In the particular case wherein the first slab is at rest in the lab frame ($v_1 = 0$), it follows that:

$$F_i^{mat} \approx F_i^{tot}\left(1 - \frac{1}{n_1^2}\right). \tag{D6}$$

Thus, since by assumption $n_i \gg 1$, the two friction forces do not differ appreciably.

It is useful to find explicitly the sign of $p_{ps,i}$. To do this, we use another result of Ref. [17, Eq. (C3)] and Eqs. (D1)-(D2) to write:

$$\frac{p_{ps,i}}{V_i} = \left(\frac{p_{ps,i}}{V_i}\right)^{co} = \frac{1}{v_{ph,i}^{co}}\left(1 - \frac{v_{ph,i}^{co} v_{g,i}^{co}}{c^2}\right)\left(\frac{E_{s,i}}{V_i}\right)^{co} \tag{D7}$$

where $V_i$ is the volume of the body, and as before the superscript $co$ indicates that a given quantity is calculated in the co-moving frame of the $i$-th body. It is simple to prove that the guided modes in our non-dispersive waveguides satisfy $|v_{ph,i}^{co} v_{g,i}^{co}| < c^2 / n_i^2$. Hence, because in the co-moving frame $E_{s,i}^{co} > 0$, we conclude that:

$$\text{sgn}(p_{ps,i}) = \text{sgn}(v_{ph,i}^{co}). \tag{D8}$$

Finally, from the selection rules (18) it follows that:

$$v_{ph,1}^{co} + v_1 = v_{ph,2}^{co} + v_2, \qquad \text{with } v_{ph,1}^{co} v_{ph,2}^{co} < 0. \tag{D9}$$

This implies that $\text{sgn}(v_{ph,i}^{co}) = -\text{sgn}(v_i - v_j)$ where $j=1$ if $i=2$, and $j=2$ if $i=1$. Using this property in Eq. (D8) we obtain Eq. (21).